\documentclass[a4paper,11pt]{article}
\pdfoutput=1

\usepackage{jheppub}
\usepackage{amssymb}
\usepackage{amsmath}
\usepackage{mathdots}
\usepackage{verbatim}
\usepackage[utf8]{inputenc}
\usepackage[usenames,dvipsnames,svgnames,table]{xcolor}
\usepackage[all]{xy}
\usepackage{graphicx}
\usepackage{mathrsfs}
\usepackage{hyperref}

\def\be{\begin{equation}}
\def\ee{\end{equation}}
\def\bea{\begin{eqnarray}}
\def\eea{\end{eqnarray}}
\def\bal{\begin{align}}
\def\eal{\end{align}}

\newcommand{\ffrac}[2]{\mbox{\footnotesize$\displaystyle\frac{#1}{#2}$}}

\newcommand{\ka}{k_\alpha}
\newcommand{\kb}{k_\beta}

\newcommand\bZ{\mathbb{Z}}
\newcommand\bC{\mathbb{C}}

\newcommand\cA{\mathcal{A}}

\newcommand\cH{\mathcal{H}}

\newcommand\cG{\mathcal{G}}

\newcommand\cT{\mathcal{T}}

\newcommand{\nnm}{\nonumber }

\newcommand{\tr}{\mathrm{tr}}

\newcommand\End{\text{End}}

\newcommand{\UH}{\mathcal{G}}
\newcommand{\one}{\boldsymbol{1}}
\newcommand{\tensor}{\otimes}

\newcommand{\comod}{\boldsymbol{a}}
\newcommand{\pivot}{\boldsymbol{g}}
\newcommand{\SSS}{\mathscr{S}}
\newcommand{\TTT}{\mathscr{T}}

\newcommand{\SSSZ}{\mathscr{S}_{Z}}

\newcommand{\TTTZ}{\mathscr{T}_{Z}}
\newcommand{\toyalg}{A_p}
\newcommand{\ribbon}{\boldsymbol{v}}
\newcommand{\coint}{\boldsymbol{c}}
\newcommand{\str}{\mathrm{str}}
\newcommand{\qtr}{\mathrm{str}_q}
\newcommand{\DA}{D_{(\cA)}}
\newcommand{\RA}{\mathfrak{R}_{(\cA)}}
\def\GL{\textit{GL}(1|1)}
\def\gl{\textit{gl}(1|1)}
\def\Ugl{\textit{U}_q\,\gl}
\def\UBgl{\overline{U}_q\,gl(1|1)}
\title{Combinatorial Quantisation of\\ $GL(1|1)$ Chern-Simons Theory I: The Torus}

\author[a]{{\color{black}Nezhla Aghaei\,}}
\author[b]{{\color{black}Azat M.~Gainutdinov\,}}
\author[c]{{\color{black}Michal Pawelkiewicz\,}}
\author[d,e]{{\color{black}Volker Schomerus\,}}

\affiliation[a] {Albert Einstein Center for Fundamental Physics, Institute for Theoretical Physics, University of Bern, Sidlerstrasse 5, Bern, ch-3012, Switzerland.}
\affiliation[b]
{Institut Denis Poisson, CNRS, Universit\'e de Tours, Universit\'e d'Orl\'eans, Parc de Grammont, 37200 Tours, France.}
\affiliation[c]{ Institut de Physique Theorique, CEA Saclay, 91191 Gif Sur Yvette, France.}
\affiliation[d]
{DESY, Theory Group, Notkestrasse 85, Building 2a, 22607 Hamburg, Germany.}
\affiliation[e]
{Department of Mathematics, University of Hamburg, Bundesstrasse 55, 20146 Hamburg, Germany.}

\emailAdd{aghaei@itp.unibe.ch,~azat.gainutdinov@lmpt.univ-tours.fr,\\~michal.pawelkiewicz@ipht.fr,~volker.schomerus@desy.de}

\abstract{Chern-Simons Theories with gauge super-groups appear naturally in string theory and they possess
interesting applications in mathematics, e.g.\ for the construction of knot and link invariants. This paper is the first in a series where we propose a
new quantisation scheme for such super-group Chern-Simons theories on 3-manifolds of the form $\Sigma \times \mathbb{R}$.
It is based on a simplicial decomposition of an $n$-punctured Riemann surface $\Sigma=\Sigma_{g,n}$ of genus
$g$ and allows to construct observables of the quantum theory for any $g$ and $n$ from basic building
blocks, most importantly the so-called monodromy algebra. In this paper we restrict to the torus case,
i.e.\ we assume that $\Sigma = \mathbb{T}^2$, and to the gauge super-group $G=GL(1|1)$. We construct  the corresponding space of quantum states
for the integer  level $k$ Chern-Simons theory along with an explicit representation of the modular group
$SL(2,\mathbb{Z})$ on these states. The latter is shown to be equivalent to the Lyubachenko-Majid action on the
centre of a restricted version of the quantised universal enveloping algebra of the Lie super-algebra $gl(1|1)$ at the primitive $k$-th root of unity.}

\begin{document}
\begin{flushright}
DESY 18-203\\
ZMP-HH/18-19\\
Hamburger Beitr\"age zur Mathematik 750
\end{flushright}
\vskip 0em

\maketitle
\flushbottom

\newpage

%%%%%%%%%%%%%%

\section{Introduction}
~~~Chern-Simons theories and their quantisation are an important research topic with many links
in particular to mathematics, such as the theory of 3-manifold invariants and knot theory.
Their role in this context was first developed in the seminal paper \cite{Witten:1988hf} and
then further explored through much subsequent work. Chern-Simons theories also play an
important role in physics. They provide key examples of topological field theories and
thereby are relevant for topological phases of matter and in particular for quantum Hall
fluids, see e.g.\ \cite{Zee:1996fe,Wen:1995qn,Witten:2015aoa,Tong:2016kpv} and many
references therein.

Most of the past research and applications have been developed for Chern-Simons theories
in which the gauge group is an ordinary (Lie) group. The generalisation to gauge supergroups,
that is also the subject of this work, has received limited attention in the past, see e.g.\
\cite{Rozansky:1992rx,Rozansky:1992td,Rozansky:1992zt,Mikhaylov:2014aoa,Mikhaylov:2015qik}.
There exist various motivations, both from physics and from mathematics,
to consider Chern-Simons theories in which the gauge connection takes values in a Lie
superalgebra. In particular, these models appear in the context of brane constructions.
As observed in \cite{Witten:2011zz}, Chern-Simons theories can emerge by topological
twisting from the intersection of $N$ D3 and NS5 branes in
10-dimensional type IIB superstring theory. Three of the four extended directions of
the D3 branes are assumed to extend along the NS5 branes while the
forth direction runs along one of the transverse coordinates $x$. The NS5
branes split this transverse direction $x$ into two disconnected parts and if we split
our stack of $N = n+m$ D3 branes into $n$ that extend to the left and $m$
extending to the right of the NS5 branes, then the topologically twisted
effective theory on the 3-dimensional intersection was shown to be
Chern-Simons theory with gauge supergroup $U(n|m)$ \cite{Witten:2011zz,Mikhaylov:2014aoa}.
The level $k$ of the Chern-Simons theory is determined by the complexified string coupling,
see \cite{Mikhaylov:2015qik} for details and references. The brane construction we sketched
here is closely related to the realisation of Chern-Simons theory through a Kapustin-Witten
topological twist~\cite{Kapustin:2006pk} of 4-dimensional $N=4$ supersymmetric Yang-Mills
theory. The latter arises as the low energy effective field theory on a stack of D3 branes.
Related constructions of Chern-Simons theories with gauge supergroup were also explored
in~\cite{Kapustin:2009cd}.

On the more mathematical side, Chern-Simons theory possesses the relation with
invariants of knots/links and 3-manifolds. If the gauge group is $G =\textit{SU}(2)$,
for example, expectation values of Wilson lines in the fundamental
representation give rise to the famous Jones polynomial. For other groups and representations
one obtains other classes of polynomials that have also been explored extensively. Knot
invariants for Lie supergroups have not been explored as much, see however
\cite{Rozansky:1992td, Rozansky:1992zt} and more recent developments in~\cite{Geer,AGeer}.
If the Wilson line operators are evaluated
in maximally atypical representations of the gauge supergroup,
the expectation values of Wilson lines turn out to be identical to the ones for a
cohomologically reduced bosonic theory, see e.g.~\cite{Candu:2010yg}. In the case the representations
are not maximally atypical, on the other hand, one expects some new invariants. Such representations
possess
zero super-dimension which causes sever problems when one attempts to
extend the usual constructions for bosonic (or purely even) gauge groups. It is one of the motivations
of our program to develop a systematic route towards such generalizations that work for
arbitrary supergroups and representations.

 We should also mention here that a  way to overcome the problem of vanishing dimensions was already proposed in~\cite{GPMT} where one uses so-called  re-normalized or modified dimensions that have nice topological properties generalising those of Reshetikhin--Turaev type. This more categorical approach has launched an avalanche of results~\cite{GKPM1,CGP, BCGP, GR, BBGe, RGP, BBG, AGeer} in a  direction related to our (though not quite directly).
We however do not follow this rather abstract route and instead use  a combinatorial approach based on graph algebras that is inspired by lattice gauge theory.

\smallskip
Another important aspect of Chern-Simons theories is their intimate relation with
2-di\-men\-si\-on\-al Wess--Zumino--Novikov--Witten  conformal field theories. 
According to common folklore, the state space of Chern-Simons theory on 3-manifold $M$ of the
form  $M = \Sigma \otimes \mathbb{R}$  with an $n$-punctured Riemann surface $\Sigma = \Sigma_{g,n}$ of genus $g$ coincides with the space
of conformal $n$-point blocks of the WZNW theory.
 For gauge supergroups, however, the relation  has not been explored well enough.
While WZNW models for gauge supergroups have been
constructed systematically~\cite{Quella:2007hr}, at least on surfaces of genus $g=0$, the state spaces of
associated Chern-Simons theories on $M = \Sigma \otimes \mathbb{R}$ were only
constructed for a few supergroups and surfaces~$\Sigma$, see in particular
\cite{Mikhaylov:2015qik} for an extensive discussion of the $\GL$ Chern-Simons theory
for a surface~$\Sigma$ of genus $g=1$. At least in this special case it was shown that
the state spaces coincide with the spaces of conformal blocks, just as expected. Through
the approach we develop below we  recover the same state space as in~\cite{Mikhaylov:2015qik}, 
but in a way that makes the generalisation to arbitrary  supergroups and surfaces 
$\Sigma$ of any genus rather straightforward. To lay the foundations for such an extension is 
indeed one of the main goals of this work. 
\smallskip

In order to do so, we  extend the combinatorial approach to the Hamiltonian
quantisation of Chern-Simons theory that was first developed in a series of papers
\cite{Alekseev:1994pa,Alekseev:1994au,Buffenoir:1994fh,Alekseev:1995rn}, and then consequently axiomatized in~\cite{MW}.
It applies to cases in which the
underlying 3-manifold $M = \Sigma \times \mathbb{R}$ splits into a spacial 2-dimensional
Riemann surface $\Sigma$ and a time direction $\mathbb{R}$. The classical phase space of
this theory is provided by the space of all gauge fields on $\Sigma$ modulo gauge
transformations. The idea of the combinatorial quantisation is to replace the continuous
space $\Sigma$ through a lattice (simplicial decomposition). While for most gauge theories
such a lattice discretisation is only an approximation, for Chern-Simons theories it is
exact due to the topological nature, at least as long as the lattice properly encodes the
topology of the underlying surface $\Sigma$. In lattice gauge theory, the group valued
holonomies of the gauge fields along the links of the lattice describe field configurations
and gauge transformations act on these holonomies at the vertices. In the classical theory
the space of such field configurations comes equipped with a Poisson bracket that respects
the gauge transformations. The combinatorial quantisation developed in \cite{Alekseev:1994pa,
Alekseev:1994au}  is achieved by replacing the algebra of functions on the link through a
$q$-deformed algebra with a deformation parameter~$q$ that is determined by the level~$k$ of the Chern-Simons theory. It can be shown that the algebra of gauge invariant
observables depends only on the underlying surface, not on the lattice discretisation.
Therefore it is possible to work with one canonical lattice, one for each surface~$\Sigma$. For an $n$-punctured Riemann surface of genus $g$, this canonical lattice
has  $2g+n$ links and a single  vertex. 
The quantum ``graph" algebra corresponding to such a lattice is made out of elementary blocks -- monodromy algebras for each closed link 
-- where the algebraic relations between different cycles elements are encoded by the quantum $R$-matrix.

These graph algebras had
a reincarnation recently within the context of factorisation homology, a notion that
was originally introduced by Beilinson and Drinfeld \cite{Beilinson} as an abstraction
from chiral conformal field and then extended to a topological setting in \cite{Lurie,
	Ayala:2015,Ayala:2017}. In \cite{Ben-Zvi:2015jua} these general concepts were made
explicit for 2-dimensional surfaces and the resulting algebras were found to agree
with those that were introduced in \cite{Alekseev:1994pa,Alekseev:1994au}.

The algebra of observables of the Chern-Simons theory carries an action of the
mapping class (or Teichm\"uller) group of the underlying surface $\Sigma$, i.e.\
of the group of orientation preserving homeomorphisms $\textit{Homeo}^+(\Sigma)$
of the surface $\Sigma$ divided by its identity component $\textit{Homeo}^+_0(\Sigma)$.
The latter consists of homeomorphisms that can be smoothly deformed to the identity.
This group is generated by so-called Dehn twists. These are special homeomorphisms
that are associated to non-contractible cycles $\gamma$ of $\Sigma$. They amount to
cutting $\Sigma$ along $\gamma$, rotating one of the resulting boundary circles by
$2\pi$ and then gluing the circles back together. In the special case of a torus, i.e.\
a Riemann surface $\Sigma_{1,0}$ this mapping class group is given by the modular group
$SL(2,\mathbb{Z})$. As usual in quantum physics this action of the mapping class group
on observables lifts to a projective action on the space of states. For Chern-Simons
theories with bosonic gauge groups $G$, the latter was worked out in \cite{Alekseev:1995rn,Alekseev:1996ns}
and it was shown to agree with the Reshetikhin-Turaev representation of the mapping
class group \cite{Reshetikhin:1990pr,Reshetikhin:1991tc,Turaevbook}.
 Let us note that this representation is intimately
related to knot and 3-manifold invariants \cite{Matveev:1994}. The relation is based on
the representation of 3-manifolds through Heegaard splitting into two handlebodies of
genus $g$. In particular, we can take a closed 3-sphere $S^3$ and remove a handlebody
$H_3$ from it. By definition, the boundary $\partial H_3$ of the 3-manifold $H_3$ is a
Riemann surface $\Sigma = \partial H_3$. Gluing this handlebody $H_3$ back into its
complement $S^3\setminus H_3$ with a non-trivial element from the mapping class group
of the surface $\Sigma$ one obtains some 3-manifold $M$. The resulting relation between
3-manifolds and elements of the mapping class group may be employed to build 3-manifold
invariants from representations of the mapping class group \cite{Kohno:1992hv}.
There exists another widely known
representation of 3-manifolds through Dehn surgery on a (framed) knot or link complement
which assigns 3-manifolds to framed links. When combined with the previous construction
one also obtains a map from elements of the mapping class group to framed links, see
\cite{Matveev:1994} for an explicit construction. Hence, representations of the mapping
class group are intimately related with invariants of 3-manifolds and of links. This may
explain our focus on the mapping class group and its representations.%
\medskip

The main goal of this paper is to discuss the quantisation of Chern-Simons theory for one
of simplest Lie supergroups, namely the supergroup $\GL$. While this will allow us to be
extremely explicit, the supergroup $\GL$ is sufficiently non-trivial to provide a prototypical
example, at least for Chern-Simons theories with gauge supergroup of type I. As we mentioned above, we expect
 an intimate relation between Chern-Simons theory and 2-dimensional
WZNW models. Supergroup  versions of the latter were studied extensively, see \cite{Quella:2013oda}
for a review. The first complete solution of the $\GL$ model was worked out in \cite{Schomerus:2005bf}.
This solution was then generalized in several steps to type~I supergroups \cite{Gotz:2006qp,Quella:2007hr}. In
the end it turned out that all the crucial elements of the theory were already visible
in the $\GL$ example, see also \cite{Quella:2013oda}.

In this work we describe a first step of a longer programme which aims at the
construction of Chern-Simons theory at both integer and non-integer levels and  for arbitrary gauge supergroups on a
manifold $M = \Sigma \otimes \mathbb{R}$ with a Riemann surface $\Sigma$ of
any genus and any number of punctures.  Compared to our goal, the actual constructions and
results we shall describe below may seem rather modest at first. In fact, we
shall focus on the Lie supergroup $\GL$, a surface $\Sigma$ of genus $g=1$
and (odd) integer level $k$. Overcoming all our restrictive assumptions is actually
less of an issue than it may naively appear. As we mentioned before, we do not
expect the extension to other supergroups  to create any new problems.
Furthermore, the combinatorial quantisation we explore here is ideally suited to
address surfaces of higher genus. The restriction to integer level may actually
seem the most problematic since representations of the modular group $\textit{SL}
(2,\mathbb{Z})$ associated with $\Ugl$ for generic~$q$ are not known so far.
Nevertheless, we will construct such representations within our approach in
a forthcoming paper.

In the next section we extend the combinatorial quantisation developed in
\cite{Alekseev:1994pa,Alekseev:1994au,VS} for semisimple (modular) Hopf algebras to not necessarily semisimple  super Hopf algebras of finite dimension, at least
for the torus $\Sigma = \mathbb{T}^2$. In this case,
the associated mapping class group coincides with the modular group $SL(2,\mathbb{Z})$.
We  describe two different actions of the latter. The first one is the action
on observables and states of Chern-Simons theory on $M = \mathbb{T}^2 \times \mathbb{R}$.
Our construction follows the general procedure in \cite{Alekseev:1995rn} and generalizes
the latter to finite-dimensional ribbon and factorisable (i.e.\ with non-degenerate monodromy) super Hopf algebras.  The second action of
the modular group we shall describe is the action on the centre of such super Hopf algebras
 introduced by Lyubachenko and Majid~\cite{Lyubashenko:1994ma}. We believe that the two actions are (projectively) equivalent in cases
where they are both well defined, see our explicit conjecture in Section~\ref{sec:conj}, but we check this claim only for the
case of $\GL$.

In Section~\ref{sec:toy}, we illustrate the general construction of quantisation at a simple (bosonic)
example, namely where the lattice gauge group is played by the group algebra of a finite cyclic group. Section~\ref{sec:4} contains
our main example which is relevant for Chern-Simons theory with the gauge supergroup
$\GL$ at integer odd level $k$. There we  introduce the restricted version of the
deformed universal enveloping algebra $\Ugl$, denoted by $\UBgl$, where the deformation
parameter $q$ satisfies $q^p = 1$ for $p$ odd integer. This is a finite-dimensional ribbon factorisable Hopf algebra. The connection to the level is simple: $p=k$ (it would be however shifted by the dual Coxeter number for other supergroups). In this case, we then describe
the combinatorial approach to the Hamiltonian quantisation of Chern-Simons theory
on the manifold $M = \mathbb{T}^2\times \mathbb{R}$. Special attention is paid to the action of
the modular group on observables and states. We construct this action  in all detail and verify that we
obtain a representation of the modular group indeed,  and in Section~\ref{sec:handle_mcg_gl11} we finally  compare this representation with the one obtained in~\cite{Mikhaylov:2015qik} based on the brane construction discussed above. Next in Section~\ref{sec:LM_mcg_gl11}, we discuss the (projective) action of the
modular group on the centre of $\UBgl$ that is defined following
Lyubachenko-Majid and show that the latter is projectively equivalent to our representation on states of the
Chern-Simons theory. In the concluding section we  outline how the results of this
work can be possibly extended to other Lie superalgebras, for surfaces of higher genus  and
beyond the cases of $q$  a root of unity.

\medskip

We should also note that at the final stage of writing this paper we became aware of very recent results~\cite{Faitg} on a related subject that proves our Conjecture from Section~\ref{sec:conj} in the purely even case.

\section{Torus observables for super Hopf algebras}
\label{Sec:2} 

In this first section we provide background material and outline the main constructions
and results. These will be illustrated through explicit examples in later sections. Our
discussion starts with a short review of ribbon super Hopf algebras. Then we turn to the
construction of monodromies and handle algebras within the framework of combinatorial
quantization that was developed for bosonic gauge groups in \cite{Alekseev:1994pa,Alekseev:1994au,Buffenoir:1994fh,Fock:1998nu}. We extend these algebras to allow for gauge
supergroups where the underlying super Hopf algebra comes from (restricted) deformed universal enveloping of the corresponding Lie superalgebra. The associated spaces of Chern-Simons states and an action of the modular
group on these states are discussed in Sections~\ref{gauge} and \ref{sec:SL2Z-handle}, following and extending the semisimple cases~\cite{Alekseev:1995rn} when necessary. We conjecture that the
resulting representation of the modular group is (projectively) equivalent to the Lyubashenko-Majid
action on the center of the underlying super Hopf algebra. For convenience of the
reader we review the latter in Section~\ref{sec:SL2Z-centre}. The conjectured equivalence between
representations is not proven in general, but in our two key examples to be discussed in Sections~\ref{sec:toy} and~\ref{sec:4}.

\subsection{Conventions on super Hopf algebras}\label{section.2.1}
In this part we recall some basics about $\mathbb{Z}_2$-graded ribbon Hopf algebras over
$\mathbb{C}$. We begin with $\mathbb{Z}_2$-graded algebras, and then recall useful identities
in the theory of integrals, and define $\mathbb{Z}_2$-graded ribbon Hopf algebras.

A $\mathbb{Z}_2$-graded algebra  $\mathcal{G}$ over the field of complex numbers $\mathbb{C}$ is a complex
$\mathbb{Z}_2$-graded vector space equipped with a multiplication map $m\colon\; \mathcal{G}\otimes \mathcal{G} \to \mathcal{G}$ and a unit map $\eta \colon\; \mathbb{C} \to \mathcal{G}$ which are $\mathbb{C}$-linear and respect
the grading. In other words,  $\mathcal{G}$  decomposes into two subspaces $\mathcal{G}=\mathcal{G}_0\oplus\mathcal{G}_1$,
called even and odd respectively, on which the multiplication acts as follows
\begin{align*}
	m\colon\; \mathcal{G}_i \otimes \mathcal{G}_j \to \mathcal{G}_{i+j},
\end{align*}
where the index is taken modulo 2. The grade of the element $a\in\mathcal{G}_i$ is defined as $|a|=i$, and we call it
\textit{even} if $|a|=0$ and \textit{odd} otherwise. The multiplication and the unit satisfy the standard algebra axioms,
including associativity. Following physics conventions we will also refer to this structure as a \textit{superalgebra}.

Having two super algebras $\mathcal{G}$ and $\mathcal{H}$, we define  $\mathbb{Z}_2$-graded algebra structure on the
tensor product $\cG\tensor \cH$ by
\begin{align}\label{grading}
	(a_1 \otimes b_1)\cdot(a_2 \otimes b_2) = (-1)^{|b_1||a_2|} a_1 a_2 \otimes b_1 b_2,
	\qquad a_1, a_2\in\cG,\; b_1,b_2\in \cH\ .
\end{align}
To define a $\mathbb{Z}_2$-graded  Hopf algebra, we also require the co-product $\Delta\colon \; \mathcal{G}\to\mathcal{G}
\otimes \mathcal{G}$  and  the co-unit $\epsilon \colon\; \mathcal{G}\to\mathbb{C}$ maps to exist. Both of them should be
$\mathbb{Z}_2$-graded algebra homomorphisms, where $\bC$ is purely even, and they are assumed to satisfy the following
co-associativity and co-unit axioms:
\begin{align}
	(\Delta\otimes id)\circ \Delta &= \Delta\circ (id \otimes \Delta),\\
	(\epsilon \otimes id)\circ \Delta &= id = (id \otimes \epsilon)\circ\Delta.
\end{align}
Let us also introduce the opposite co-product
\be
\Delta^{\mathrm{op}}  := \tau\circ \Delta,
\ee
where we used the flip map of super vector spaces $\tau\colon \; \mathcal{G}\otimes \mathcal{G} \to \mathcal{G}\otimes
\mathcal{G}$ defined on homogeneous elements as
\be\label{eq:sflip}
\tau(a \otimes b)=(-1)^{|a||b|} b\otimes a \ .
\ee
Further,  we require a grade preserving map $S \colon\;  \mathcal{G}\to\mathcal{G}$, called an antipode, which is  an
algebra anti-homomorphism, that is
\begin{equation}
	S(ab) = (-1)^{|a||b|} S(b)S(a) , \qquad a,b\in\mathcal{G},
\end{equation}
and a co-algebra anti-homomorphism, that is
\be
\Delta \circ S = (S\tensor S)\circ \Delta^{\mathrm{op}}.
\ee
In addition, it also satisfies
\begin{align}
	m \circ (id\otimes S)\circ \Delta = m\circ (S\otimes id)\circ \Delta = \eta\circ\epsilon .
\end{align}
Finally, a $\mathbb{Z}_2$-graded algebra $\mathcal{G}$ equipped with a co-product $\Delta$, a co-unit $\epsilon$ and an
antipode~$S$ is called  \textit{$\mathbb{Z}_2$-graded} or \textit{super Hopf algebra}, and will be denoted by the same
symbol $\mathcal{G}$ as the underlying algebra in the following.

A super Hopf algebra is \textit{quasi-triangular} if its tensor square admits an invertible element called
\textit{universal $R$-matrix} $R \in (\mathcal{G}\otimes \mathcal{G})_0$ satisfying the following relations
\begin{align}
	& R \cdot \Delta(x)= \Delta^{\mathrm{op}}(x) \cdot R, \qquad x\in\mathcal{G}, \notag\\
	& (id \otimes \Delta)(R) = R_{13} \cdot R_{12}, \label{eq:R-2}\\
	& (\Delta \otimes id)(R) = R_{13} \cdot R_{23} , \notag
\end{align}
where we set
\begin{align}\label{eq:R-indices}
	R_{12} &= R\otimes 1, \quad
	R_{23} = 1\otimes R, \\
	R_{13} &= (id\otimes\tau)(R\otimes 1) ,\notag
\end{align}
with $\tau$ as defined in equation~\eqref{eq:sflip}.

One can define a monodromy matrix $M\in\mathcal{G}\otimes \mathcal{G}$ using the universal $R$-matrix
\begin{align}\label{universal}
	M = R' \cdot R,
\end{align}
where
$$
R'=R_{21}=\tau(R)
$$
and we keep using `$\cdot$' notation to emphasise that the product is as in equation~\eqref{grading}.
We will call a monodromy matrix $M$ \textit{non-degenerate} if it can be expanded
\be\label{eq:M-nondeg}
M=\sum_{i=1}^{\dim(\cG)} f_i\tensor g_i
\ee
with $\{f_i\}$ and $\{g_i\}$ being two bases in $\cG$. If such an expansion of $M$ exists, we
call $\cG$ \textit{factorisable}.

A $\mathbb{Z}_2$-graded  Hopf algebra $\mathcal{G}$ is called \textit{ribbon} if it admits  a so-called
ribbon element $\ribbon\in\mathcal{G}$, which is  an even central element  satisfying
\be\label{eq:rib-ax}
M \cdot \Delta(\ribbon) = \ribbon\otimes \ribbon, \qquad
S(\ribbon) = \ribbon.
\ee
We note that in the case of semisimple ribbon and factorisable $\cG$, the name ``modular Hopf algebra"
is also used because representations of $\cG$ provide then a modular category.

In a ribbon super Hopf algebra, we have the identities
\be
\ribbon^2 = \boldsymbol{u} S(\boldsymbol{u}), \qquad
\epsilon(\ribbon) = 1 ,
\ee
where we used the so-called Drinfeld  element
\be\label{eq:u-el}
\boldsymbol{u} = m\circ(S\otimes id) (R')\ .
\ee
One can find an explicit expression for the ribbon element from the properties of a right integral in
the manner described below.

\subsection{Integrals and co-integrals}\label{section.2.2}
We now review standard facts from the theory of integrals~\cite{Ra-book} for a Hopf algebra. We will use
the same theory in our super algebra setting, as super Hopf algebras are normal Hopf algebras too.

A linear form $\mu\in\mathcal{G}^*$ will be called a \textit{right integral of $\cG$} if it satisfies
\begin{equation}\label{integral}
	(\mu \otimes id) \circ\Delta(x) = \mu(x) \one,
\end{equation}
for all $x\in\mathcal{G}$. Similarly, one can  define a left integral, with $\mu$ hitting the second
tensor factor instead. If an integral exists it is unique up to a scalar. Moreover, it is known that a
finite-dimensional Hopf algebra always allows such integrals~\cite{LarsonSweedler}. However in general
a right integral  does not have to coincide with a left one. Such a deviation of a right integral from
being a left one is measured by a group-like element $\comod$ called \textit{co-modulus}: \footnote{It
	is called \textit{the distinguished group-like} element in~\cite{Ra-book}.}
\begin{equation}\label{left_integral}
	(id \otimes \mu) \circ \Delta(x) = \mu(x) \comod.
\end{equation}

\textit{A right co-integral of $\UH$} is an element $\coint\in \UH$ such that
\be\label{eq:coint-def}
\coint x=\epsilon(x) \coint\ , \qquad x\in \UH\ .
\ee
We note that this notion is actually dual to the notion of the integral: under a canonical identification
between $\cG$ and $\cG^{**}$, the element $\coint$ defines a right integral of $\cG^*$. We can likewise
define the left co-integral using instead the left multiplication by $x$. Non-trivial right and left
co-integrals are unique up to scalar~\cite{LarsonSweedler}. A Hopf algebra is called {\em unimodular} if
its right   co-integral is also left.
\medskip

In the case when we have a universal $R$-matrix and corresponding $M$-matrix is  non-degenerate, the
integral can be normalised by
\begin{align}\label{normalisation}
	(\mu\otimes \mu)(M) = 1.
\end{align}
This will be the case for our examples below. From now on, we will consider only finite-dimensional
quasi-triangular Hopf algebras with a non-degenerate monodromy matrix.

If the co-modulus can be expressed as a square of a group-like element in $\cG$, i.e.
\begin{align}\label{balancing}
	\comod = \pivot^2 ,
\end{align}
then such an element $\pivot$ satisfies $S^2(x)=\pivot x\pivot^{-1}$, for $x\in\cG$, and  it is called a
\textit{balancing element}.

The balancing element is important for two reasons. First, it provides us the ribbon element
\begin{equation}\label{ribbon}
	\ribbon = \pivot^{-1} \boldsymbol{u} ,
\end{equation}
where we recall the Drinfeld element~\eqref{eq:u-el}. This is a concrete formula for the ribbon
element that will be used in the following sections.

Secondly, the balancing element provides us with a notion of \textit{quantum trace} over a
representation $\pi\colon \cG \to \End(V)$,
\be\label{eq:q-tr}
\qtr\bigl(\pi(x)\bigr) := \str\bigl(\pi(\pivot^{-1} x)\bigr)\ ,\qquad x\in\cG\ ,
\ee
where $\str(-)=\tr(\omega(-))$ is the supertrace with $\omega$ the parity map. The quantum trace over a representation $\pi$ of $\cG$ can be used to produce a central element of $\cG$:
\be\label{eq:z-pi}
z_\pi :=   \big((\text{str}_q \circ \pi) \otimes id\big)(M)\; \in \; Z(\cG),
\ee
see e.g.~\cite{[Drinfeld]} for non-graded case.
Though in general not all central elements of $\cG$ can be produced this way: the map $\mathrm{Rep}\, \cG \to Z(\cG)$ defined in~\eqref{eq:z-pi} is surjective if only if the algebra $\cG$ is semisimple.

\subsection{Reconstruction of $\cG$}\label{sec:monodromy}
Let $\cG$ be a finite-dimensional factorisable (super) Hopf algebra. We recall that the monodromy matrix
$M$ from \eqref{universal} is an element in $\mathcal{G}\otimes\mathcal{G}$ which can be expanded as
in \eqref{eq:M-nondeg} where the two sets of elements $f_i$ and $g_i$ both provide a basis of $\cG$.
The algebra $\cG$ can be reconstructed from such non-degenerate $M$. Indeed, the linear map
\be
\cG^*\to \cG, \qquad f\mapsto (f\otimes id) (M)
\ee
is an isomorphism if and only if $M$ is non-degenerate, e.g.\ one can run here over $f$ being the dual
elements $f^*_i$ to the basis $f_i$ to recover all basis elements $g_i\in\cG$.

The reconstruction of $\cG$ from $M$ can be processed also on the algebraic level.
We first recall that $M$ satisfies an ``exchange" relation:
\begin{align}\label{RMRMrelation}
	R_{21} \cdot M_{13} \cdot R_{12} \cdot M_{23} = M_{23} \cdot R_{21} \cdot M_{13}\cdot R_{12} ,
\end{align}
which follows straightforwardly from the relations~\eqref{eq:R-2}, and here we used conventions for $M_{ij}$
as in \eqref{eq:R-indices}. We then  think of the above relation as a set of $\text{dim}(\mathcal{G})^2$
(anti-)commutation relations for the elements in the third tensor factor, each relation  corresponding to a
basis element of  $\mathcal{G}\otimes\mathcal{G}$. In terms of the basis expansions \eqref{eq:M-nondeg} this
means the following: Let us introduce $R= \sum_{i,j} a_i\tensor b_j$ which serve us as ``structure constants"
matrix. Then the above equation  can be written in components as follows
\begin{align}\begin{split}\label{eq:monodromy-eq-basis}
		&\sum_{i,j,k,l,m,n=1}^{\dim(\cG)} (-1)^{|a_i|(|b_j|+|f_k|+|a_l|) + |g_k|(|a_l|+|b_m|+|f_n|)} b_j f_k a_l \otimes a_i b_m f_n \otimes g_k g_n = \\
		&=\sum_{i,j,k,l,m,n=1}^{\dim(\cG)} (-1)^{|a_i|(|b_j|+|f_k|+|a_l|) + |f_n||b_m|+|f_k|(|a_l|+|b_m|)} b_j f_k a_l \otimes f_n a_i b_m \otimes g_n g_k,
\end{split}\end{align}
where we simplified the minus signs by taking into account that the monodromy matrix is an even element in
$\cG\otimes\cG$, i.e. $|f_i|=|g_i|$. It is clear that using the (anti-)commutation relations of $\cG$ one can
arrange the elements on the second tensor factor of the right hand side of the above equation in such a way
that they agree with those on the second tensor factor on the left, and by equating the corresponding terms
we thus obtain defining  relations for the third tensor factor in terms of  the basis elements $g_k$. In
fact, using equation \eqref{RMRMrelation} one can reconstruct the relations of the initial  algebra
$\mathcal{G}$ without knowing the commutation relations of the elements on the third tensor factor. We will
pursue a similar treatment for an algebra which we will define as the handle algebra $\mathcal{T}$.

\subsection{Handle algebra $\mathcal{T}$ and its Fock representation}\label{section.2.4}

In this section, we  describe how to define a so-called handle algebra
for a given ribbon super Hopf algebra. We will see that certain elements of the handle algebra
give a realisation of the $SL(2,\mathbb{Z})$ group, i.e.\ the mapping class group of the torus,
acting on its Fock-type representation.

Let $\cG$ be a finite-dimensional factorisable (super) Hopf algebra. One can  define an algebra
using the notion of  \textit{universal element}~\cite{VS}, which belongs to a tensor product of
the Hopf algebra~$\cG$ and of the algebra being defined, subject to a set of equations. We have
already encountered in Section~\ref{sec:monodromy}  an example of a universal element given by
the monodromy matrix $M$ of the Hopf algebra $\cG$, which one can regard as an element in
$\mathcal{G}\otimes\mathcal{G}$.

The handle algebra $\mathcal{T}$ is defined using a pair of universal elements $A,B\in\mathcal{G}
\otimes\mathcal{T}$ subject to the exchange relations
\begin{align}
	R_{21} \cdot A_{13} \cdot R_{12}  \cdot A_{23} &= A_{23} \cdot R_{21}\cdot  A_{13} \cdot R_{12} , \label{handlealg_firstexchangerelation}\\
	R_{21} \cdot B_{13} \cdot R_{12} \cdot B_{23} &= B_{23} \cdot R_{21} \cdot B_{13} \cdot R_{12} ,\label{handlealg_secondexchangerelation} \\
	R^{-1}_{12} \cdot A_{13} \cdot R_{12} \cdot B_{23} &= B_{23 } \cdot R_{21} \cdot A_{13}\cdot R_{12} , \label{handlealg_thirdexchangerelation}
\end{align}
which are equations in the vector space $\cG\tensor \cG \tensor \cT$. In a fixed basis in $\cT$,
these equations can be written explicitly in the same manner as \eqref{eq:monodromy-eq-basis}.
In other words,  the handle algebra $\mathcal{T}$ is generated algebraically by the images of $A$
and $B$ under the ``evaluation" map
\be
f\tensor id: \; \cG\tensor\cT\to \cT\ , \qquad \text{for} \; f\in \cG^*\ .
\ee
In particular, running over all $f\in\cG^*$  and applying this map to $A$ we recover a subalgebra
in $\cT$ isomorphic to $\cG$, as this single element $A$ satisfies the same relation~\eqref{RMRMrelation}
as the monodromy matrix $M$ does. We will thus call $A=M(a)$ the universal element or monodromy
corresponding to the $a$-cycle of the torus, and the corresponding algebra \textit{loop} or \textit{monodromy algebra}. Similarly, we call $B=M(b)$ the monodromy of the $b$-cycle.
Therefore, the universal elements $A$ and $B$ generate two subalgebras $\mathcal{T}^{(a)}$ and
$\mathcal{T}^{(b)}$ of $\mathcal{T}$, which are both isomorphic to $\mathcal{G}$, and the units
of these subalgebras coincide with the unit of~$\cT$. However, the elements from $\mathcal{T}^{(a)}$
do not commute with those from $\mathcal{T}^{(b)}$ -- the third exchange relation
\eqref{handlealg_thirdexchangerelation} provides non-trivial commutation relations between elements
of such subalgebras. Explicit examples of this construction will be provided in the next two
sections for both semisimple and non-semisimple cases.
\medskip

The handle algebra $\mathcal{T}$ has a representation (this is motivated by~\cite[Thm.\,21]{Alekseev:1995rn})
$$
D\colon \; \cT \to \End_{\bC} \,\mathfrak{R}
$$
on a finite-dimensional vector space $\mathfrak{R}$ that has the form of a Fock module which is constructed in two steps:
\begin{enumerate}
\item
Introduce a vacuum state $|0\rangle\in\mathfrak{R}$ which is by definition left invariant with respect to  the universal element $B$
associated to  the $b$-cycle
\begin{align}\label{reprelation}
	\bigl\{(id\otimes D)B\bigr\} (1 \otimes |0\rangle) = 1 \otimes |0\rangle .
\end{align}
This is an equation on elements of $\mathcal{G}\otimes \mathfrak{R}$.
The only solution of this equation is
\be\label{eq:B-act-R}
g^{(b)}  |0\rangle = \epsilon(g) |0\rangle,
\ee
where $g\in \cG$ and $g^{(b)}$ is the corresponding element in $\cT^{(b)}$ under the isomorphism $\cG\cong\cT^{(b)}$. We also note that practically we can  rewrite the above equation~\eqref{reprelation} ``component-wise" on representations $\pi$ of $\cG$ as
$$
\bigl\{(\pi\otimes D)B\bigr\} (v\tensor |0\rangle) = v\tensor |0\rangle,
$$
for all representations $\pi$ and for all vectors $v$ in the representation space of $\pi$.
This gives us a system of equations for the action of elements from the second tensor factor,
i.e.\ the algebra $\cT^{(b)}$, on the vacuum, c.f.~\cite[Eq.\,(7.3)]{Alekseev:1995rn}\footnote{We
	note that we however use a  different normalisation, which facilitates a comparison of two $SL(2,\bZ)$
	actions discussed below.}.
	
	\item
The Fock module  $\mathfrak{R}$ is then defined as a free module over $\cT^{(a)}$ generated from the vacuum~$|0\rangle$. In other words, the rest of the vectors $|f\rangle\in\mathfrak{R}$ that belong to the representation
space is obtained by applying the elements from the subalgebra $\mathcal{T}^{(a)}$ on the vacuum without imposing
extra relations, and these are
\begin{align}\label{vectors_of_fock_module}
	|f\rangle = \bigl\{(f \otimes D)A\bigr\} |0\rangle , \qquad \text{for} \; f\in\mathcal{G}^* \ .
\end{align}
As the solution for $A$ will be provided by the non-degenerate monodromy matrix $M=M(a)$, in this case the Fock module
is isomorphic to $\cG$ as a vector space. We thus see by construction that $\mathfrak{R}$ is a representation of
$\cT^{(a)}$, it is actually isomorphic to the regular representation of $\cG\cong\cT^{(a)}$. We thus only need to show that the
action of $\cT^{(b)}$ is well-defined on such vectors -- this follows from the third exchange
relation~\eqref{handlealg_thirdexchangerelation} in $\cT$ that provides commutation relations between elements in
$\cT^{(a)}$ and $\cT^{(b)}$. These are obtained from equations analogous to~\eqref{eq:monodromy-eq-basis} and they are of the form:\footnote{Formally, we have $\dim(\cG)^2$ number of such relations, though not necessarily all of them are  algebraically independent.}
\be\label{eq:rels-A-B}
\sum_{k,n=1}^{\dim(\cG)} f_{k,n}\, g^{(a)}_k \cdot g^{(b)}_n =  \sum_{k,n=1}^{\dim(\cG)} \tilde{f}_{k,n} \, g^{(b)}_n \cdot g^{(a)}_k \ ,
\ee
where $f_{k,n}$ and $\tilde{f}_{k,n}$ are complex numbers (possibly zero), and $g^{(a)}_k$ and  $g^{(b)}_n$ are elements in the basis expansions~\eqref{eq:M-nondeg} corresponding to $M(a)$ and $M(b)$ respectively.
Under the action of $\cT^{(b)}$ on the free $\cT^{(a)}$-module $\mathfrak{R}$, we  can always use~\eqref{eq:rels-A-B} to pass elements of $\cT^{(b)}$ through those of
$\cT^{(a)}$ until they reach the vacuum where we already fixed the action via~\eqref{eq:B-act-R}.
\end{enumerate}

We recall that in the case when $\cG$ is semisimple, it is known that the handle algebra~$\mathcal{T}$ can be also constructed as the Heisenberg double of the Hopf algebra $\mathcal{G}$.
We did not investigate an analogue of this in the non-semisimple case but we believe that such an isomorphism to a Heisenberg double also holds. Furthermore, due to this relation to the Heisenberg double, the handle algebra in the semisimple case admits a unique irreducible representation given by the Fock module defined above. Again, in the non-semisimple case it is an open problem of classification of representations of $\cT$ but for the analysis of $SL(2,\bZ)$ action below we will need  the Fock module only.

\subsection{Gauge invariant subalgebra and its representation}\label{gauge}
In analogy with the construction in the semisimple case~\cite{VS}, our next step in defining the $SL(2,\bZ)$ action is
to introduce the so-called gauge invariant subalgebra $\mathcal{A}$ in the handle algebra $\cT$. To this end let us first
define a so-called ``adjoint" action  of the Hopf algebra $\mathcal{G}$ on~$\mathcal{T}$. We recall that $\cT$ is generated
by elements from its  $a-$ and $b-$ cycle subalgebras $\cT^{(i)}\subset\cT$, for $i=a,b$, as was discussed in
Section~\ref{section.2.4}. Moreover,  we have an  algebra isomorphism:
\be
\kappa^{(i)}\colon\; \cG \to \cT^{(i)}\ , \quad M \mapsto M(i)
\ee
written in terms of the universal elements. We can now define the $\cG$-action of ``adjoint" type on these
subalgebras:
\begin{align}\label{action_of_cG_on_cT}
	x(f) = \sum_{(x)} (-1)^{|f| |x''|} \kappa^{(i)}(x') \cdot f \cdot \kappa^{(i)}\bigl(S(x'')\bigr) \ ,
	\qquad x\in\cG \ ,\; f\in\cT^{(i)}\ ,
\end{align}
where we used the standard Sweedler notation for co-product components
\begin{align}
	\Delta(x) = \sum_{(x)} x' \otimes x'' .
\end{align}
In particular, using the Hopf algebra axioms we have $x(1) = \epsilon(x)1$.

The action~\eqref{action_of_cG_on_cT} makes $\cT$ a module algebra over the Hopf algebra $\cG$, i.e.\
the action is compatible with the multiplication in $\cT$ in the sense that
\be\label{eq:x-fg}
x(f\cdot g) = \sum_{(x)} (-1)^{|f| |x''|} x'(f)\cdot x''(g)\ ,
\ee
where $x\in\cG$ and $f,g\in\cT$. Therefore, one can construct a smash product algebra ${\mathcal{\bar T}} =
\mathcal{T} \rtimes \mathcal{G}$ by defining the multiplication
\begin{align}\label{eq:smash-product}
	(f\otimes x)\cdot(g\otimes y) = \sum_{(x)} (-1)^{|g| |x''|} f \cdot x'(g)\otimes x'' \cdot y ,
\end{align}
where $f,g\in\cT$ and $x,y\in\cG$. We will denote the element $f\otimes x$ as $f \cdot\iota(x)$, where
$$
\iota \colon\;  \mathcal{G} \to \mathcal{\bar T}, \qquad x\mapsto 1\tensor x,
$$
is the canonical embedding map. Using \eqref{eq:smash-product} for the choice $y=1$ and $f$ equal
the unit in $\cT$, we  note the relation
\be\label{eq:G-T-rels}
\iota(x) \cdot g = \sum_{(x)} (-1)^{|g| |x''|}  x'(g)\cdot \iota(x'') \ , \qquad x\in\cG\;,\; g\in\cT\ ,
\ee
that we use below. The smash product ${\mathcal{\bar T}} = \mathcal{T} \rtimes \mathcal{G}$ can be
alternatively and equivalently defined \cite{VS, Alekseev:1995rn} using the universal elements $A$ and $B$, by the following relations
\begin{align}\label{gaugerelations1}
	\big\{(id\otimes \iota)\Delta(x)\big\} \cdot A &= A \cdot\big\{(id\otimes \iota)\Delta(x)\big\}, \\ \label{gaugerelations2}
	\big\{(id\otimes \iota)\Delta(x)\big\} \cdot B &= B\cdot \big\{(id\otimes \iota)\Delta(x)\big\},
\end{align}
where $x\in\mathcal{G}$ and the universal elements $A$ and $B$ are considered as elements of $\cG\otimes
{\bar \cT}$, i.e.\ in $\cG\otimes \cT\otimes \cG$ where we identify $A$ with $A\tensor 1 \in \cG\otimes
\cT\otimes \cG$ and $B$ with $B\tensor 1 \in \cG\otimes\cT\otimes \cG$.

Finally, the \textit{gauge invariant} subalgebra $\mathcal{A}$ of the handle algebra is defined as the
subalgebra  of $\cG$-invariant elements
\begin{align}\label{eq:cA-def}
	\mathcal{A} = \{ f \in \cT \subset \mathcal{\bar T}\; | \; \iota(x) f = (-1)^{|f||x|} f \iota(x), \; \forall\,x\in\mathcal{G} \}.
\end{align}
We note that the above definition of $\cA$ is equivalent to
$$
\cA =  \{ f \in \cT \; | \; x(f) = \epsilon(x) f, \; \forall\,x\in\mathcal{G} \}
$$
and it is clear that $x(f\cdot g) = \epsilon(x) f\cdot g$, for $f,g\in\cA$, as follows from~\eqref{eq:x-fg} and using the super Hopf algebra axioms on the co-unit. Therefore $\cA$ forms indeed a subalgebra in $\cT$.

 We show now that $\cA$ contains an important subalgebra, the one generated by the two centres  $Z(\cT^{(a)})$ and $ Z(\cT^{(b)})$.
The crucial observation here is that central elements in a Hopf algebra $H$ are invariants under the adjoint action, i.e.\ if $z\in Z(H)$ then $\sum_{(x)}x' z S(x'') = \epsilon(x)z$ for all $x\in H$. The same applies for the super Hopf algebras where the adjoint action is now defined as in~\eqref{action_of_cG_on_cT}, i.e.\ with the sign factors\footnote{However, we note that as central elements are even the signs in~\eqref{action_of_cG_on_cT} and in~\eqref{eq:cA-def} do not play a role.}. We thus have that under the $\cG$ action~\eqref{action_of_cG_on_cT} on the subalgebra $\cT^{(a)}$ the following equalities hold for all $x\in \cG$ and $z\in Z(\cT^{(a)})$:
 \be
  x(z) = \epsilon(x) z,
 \ee
 and similarly for the $b-$cycle centre $Z(\cT^{(b)})$.
We thus get that
\be\label{eq:Z-ab-in-A}
  Z(\cT^{(a)})\subsetneq \cA , \qquad Z(\cT^{(b)}) \subsetneq \cA
\ee
and all products of elements from the two centres belong to $\cA$ too. However, the two centres do not in general generate the algebra $\cA$, as it can be seen on the example of $\UBgl$ in next sections.

The Fock representation  of the handle algebra $\mathcal{T}$ can be
extended to a representation of $\mathcal{\bar T}$ and then to one of $\mathcal{A}$. For any element
$\iota(x)$, $x\in\mathcal{G}$,  we impose that on the representation space $\mathfrak{R}$ it acts in
the following way (compare with~\cite[Prop.\,12]{VS})
\begin{align}
	D(\iota(x))|0\rangle = \epsilon(x) |0\rangle \ ,
\end{align}
while the action on other vectors is obtained using the commutation relations~\eqref{eq:G-T-rels}.
Indeed, recall that any vector in the Fock module can be written as $D(g) |0\rangle$ for some
$g\in\cT^{(a)}$. Therefore we can write
\be\label{eq:D-G-rep}
D(\iota(x))\bigl(D(g)|0\rangle\bigr) = D\bigl(\iota(x) \cdot g\bigr) |0\rangle =
\sum_{(x)} (-1)^{|g| |x''|}  D\bigl(x'(g)\bigr)\cdot \epsilon(x'')  |0\rangle =  D\bigl(x(g)\bigr)|0\rangle.
\ee
Here, we first used the requirement that $D$ is an algebra map, then \eqref{eq:G-T-rels} in the
second equality, and then the Hopf algebra co-unit axiom for the last equality -- however a comment
is necessary for odd elements  $g$:  the sign factors $(-1)^{|g| |x''|}$ are actually irrelevant due
to the fact that $\epsilon$ is an even map, in particular  $\epsilon(x'')=0$ for odd $x''$, while the
sign is $+1$ for even $x''$. This finally proves~\eqref{eq:D-G-rep}, and therefore $D$ indeed
constitutes a representation of $\mathcal{\bar T}$. For brevity, we will use the same notation $D$
for both $\cT$ and $\mathcal{\bar T}$.

Of course, we can define a representation of $\cA$ as a restriction of $D$ to $\cA$. However, we need a much smaller space -- the subspace of $\cG$-invariants in $\mathfrak{R}$ that can be formally defined as
\be\label{eq:Inv-R}
\mathrm{Inv}(\mathfrak{R}) := \mathrm{Hom}_{\cG}(\mathbb{C},\mathfrak{R}).
\ee
 This subspace corresponds in~\cite{Alekseev:1995rn} to the ``flatness" restriction on $D$. In the semisimple case, such a restriction can be constructed using an appropriate projector. In the non-semisimple case, such a projector generally does not exist.  We however do not need to follow this way as the space $\mathrm{Inv}(\mathfrak{R})$ of gauge-invariant states, i.e.\ those that $D(\iota(x))|f\rangle = \epsilon(x) |f\rangle$, can be constructed directly from the (gauge-invariant) vacuum $ |0\rangle$ by applying all possible gauge-invariant operators, and we know that these are all in $\cA$.
We thus consider a ``truncation" of the representation $D$ to a representation $\DA$ of the gauge-invariant
subalgebra $\cA$ generated  from the vacuum by $\cA$:
\be\label{eq:RA-def}
\RA := D(\cA) | 0\rangle.
\ee
This clearly defines a representation of $\cA$, which is a subspace in $\mathfrak{R}$. Assume now $g\in\cA$ then $D(g)|0\rangle$ is a gauge-invariant state -- indeed, this follows from~\eqref{eq:D-G-rep} because $x(g)=\epsilon(x)g$. In other words we have shown that $\RA \subset \mathrm{Inv}(\mathfrak{R})$.
Moreover, we claim that the subspace $\RA$ contains all  gauge-invariant states, i.e.
\be\label{eq:RA-Inv}
\RA = \mathrm{Inv}(\mathfrak{R}).
\ee
 This follows from the fact that $|0\rangle$ is a cyclic vector generating  the whole module $\mathfrak{R}$ under the action of $\cT$, and similarly all the gauge-invariant states are generated from this cyclic vector by the centraliser of $\cG$, which is $\cA$ by definition. We note the importance of the \textsl{cyclic} vector: in the semisimple case, $\cA$ acts on the multiplicity space of $\cG$-invariants via an irreducible representation, and thus it would be enough to use any non-zero gauge-invariant state to produce the whole space of $\cG$-invariants via  the action of $\cA$ on it; in the non-semisimple case, the action of $\cA$ on the multiplicity space~\eqref{eq:Inv-R} is not necessarily irreducible but it is indecomposable, and thus from a gauge-invariant state we might generate only a proper subspace in $\mathrm{Inv}(\mathfrak{R})$, however from a cyclic vector the action of $\cA$  generates  the whole space of $\cG$-invariants.

 It is clear that $\RA$ contains an important subspace generated by the $a$-cycle centre:
$$
D\big(Z(\cT^{(a)})\big) | 0\rangle \subset \RA.
$$
From the action~\eqref{eq:B-act-R} of the $b$-cycle centre on the vacuum, it is also clear that the algebra  generated by both centres $Z(\cT^{(a)})$ and $Z(\cT^{(b)})$ (a subalgebra in $\cA$) generates the same subspace $D\big(Z(\cT^{(a)})\big) | 0\rangle$. We will however see in our non-semisimple example in Section~\ref{sec:4} that the gauge-invariant algebra $\cA$ is much bigger than  the algebra  generated by $Z(\cT^{(a)})$ and $Z(\cT^{(b)})$. Assume now $a\in\cA$ is an element that is not  necessarily written as a product of elements from $Z(\cT^{(a)})$ and $Z(\cT^{(b)})$. It is however can be written as, recall relations \eqref{eq:rels-A-B} and that $\cT$ has dimension $\dim(\cG)^2$,
$$
a = \sum_{k,n=1}^{\dim(\cG)} f_{k,n}\, g^{(a)}_k \cdot g^{(b)}_n \  ,
$$
for some numbers $f_{k,n}$. Applying such a general element on the vacuum and using~\eqref{eq:B-act-R}  we get
$$
a  | 0\rangle  = \sum_{k,n=1}^{\dim(\cG)} f_{k,n}\,\epsilon\bigl(g^{(b)}_n\bigr) g^{(a)}_k | 0\rangle \ ,
$$
i.e.\ we have
$$
a  | 0\rangle \; \in \; D\big(\cT^{(a)}\big)| 0\rangle\ .
$$
But we assumed  that $a\in\cA$ or $a  | 0\rangle$ is a $\cG$-invariant, and the only operators from $D\big(\cT^{(a)}\big)$ that produce $\cG$-invariants from $| 0\rangle$ are invariants  under the adjoint action, or operators  from $D\big(Z(\cT^{(a)})\big)$.
We thus conclude that
\be\label{eq:RA-Z}
\RA = D\big(Z(\cT^{(a)})\big) | 0\rangle .
\ee
In other words, as a vector space $\RA$ is isomorphic to the centre of $\cG$.

 We will use this   representation $\RA$ for our formulation of the (projective) action
of the mapping class group of the torus.

\subsection{Representation of mapping class group of the torus}\label{sec:SL2Z-handle}
In this section, we describe the realisation of the $SL(2,\mathbb{Z})$ group through elements
of the gauge-invariant subalgebra $\mathcal{A}$ of the handle algebra $\mathcal{T}$.
Then, we define our projective action of $SL(2,\mathbb{Z})$ on the subspace $\RA$ in $\mathfrak{R}$ generated by $\cA$ from the gauge-invariant vacuum. (This projective representation can be interpreted as  the space of Chern-Simons observables.)
In order to define such a representation, we first recall some facts
about the mapping class group of the torus.
\medskip

The first homotopy group $\pi_1(\mathbb{T}^2)$ of the torus is generated by the elements
$a,b$ associated to the corresponding cycles on $\mathbb{T}^2$, which are subjected to the following relation
\begin{align}\label{flatness_classical_pi1}
	b a^{-1} b^{-1} a = e .
\end{align}
This relation is interpreted as a lack of punctures or discs removed from the torus.
On the group $\pi_1(\mathbb{T}^2)$, one can define two automorphisms $\alpha$ and  $\beta$ which act as
\begin{align}\label{eq:auto-cl}
	&\alpha(a) = a , && \alpha(b) = ba, \\ \nnm
	&\beta(a) = b^{-1} a , && \beta(b) = b,
\end{align}
and they can be interpreted as Dehn twists along the $a$- and $b$-cycles. Recall that $\mathrm{Aut}\big(\pi_1(\mathbb{T}^2)\big)$ is $SL(2,\mathbb{Z})$. We can relate those
automorphisms to the standard generators $\sigma,\tau$ of $SL(2,\mathbb{Z})$ as follows
\begin{align}
	\sigma &= \alpha\circ\beta\circ\alpha = \beta\circ\alpha\circ\beta, \\
	\tau &= \alpha^{-1} .
\end{align}
It is easy to see that they satisfy the expected relations
\begin{align}
	\sigma^4 &= id
	&& (\sigma\tau)^3 = \sigma^2 .
\end{align}

The main idea now is to use a quantised version of the automorphisms $\alpha$, $\beta$.
We have seen in the previous
sections that in defining a quantum theory we associate the universal elements $A$ and $B$ to the $a$-and
$b$-cycles respectively.  In fact,
the handle algebra $\mathcal{T}$ admits a pair of automorphisms $\alpha,\beta \colon\; \mathcal{T}\to\mathcal{T}$ which realise a ``quantum"
version of the action~\eqref{eq:auto-cl}:\footnote{The appearance of the ribbon element, when
compared to the classical equations above, reflects  the quantum nature of the automorphisms
$\alpha$ and $\beta$.}
\begin{align} \label{quantumaction}
\begin{aligned} & (id\otimes\alpha)(A) = A, \\
		&(id\otimes\alpha)(B) = (\ribbon^{-1}\otimes 1) \cdot BA, \end{aligned} &&
	\begin{aligned} & (id\otimes\beta)(A) = (\ribbon\otimes 1)\cdot B^{-1} A, \\ &(id\otimes\beta)(B)
		= B, \end{aligned}
\end{align}
where   $\ribbon$ is the ribbon element of $\cG$ introduced in~\eqref{ribbon}. 
That $\alpha$ and $\beta$ are automorphisms, i.e.\ respect the relations~\eqref{handlealg_firstexchangerelation}-\eqref{handlealg_thirdexchangerelation}, is proven along the same lines as in the proof of~\cite[Lem.\,6]{VS} where the semisimplicity assumption on $\cG$ was not actually used but only the general properties of $\ribbon$ and of the universal elements $A$ and $B$ that are valid in our case too.

The automorphisms $\alpha$ and $\beta$ can be expressed as
inner automorphisms of the handle algebra, given by the adjoint actions
\be\label{automorphisms_as_adj_of_dehn_twists}
	\alpha(x) = \big({\hat v}(a)\big)^{-1} \cdot x \cdot {\hat v}(a) , \qquad \beta(x) =\big({\hat v}(b)\big)^{-1} \cdot x \cdot {\hat v}(b) , \qquad x\in\cT,
\ee
of the following elements of the handle algebra $\mathcal{T}$
\begin{align}\label{dehn_twist_quantum_eq1}
\begin{split}
	{\hat v}(a) &= (\mu\otimes id)\big((\ribbon^{-1}\otimes 1)\cdot A\big), \\
	{\hat v}(b) &= (\mu\otimes id)\big((\ribbon^{-1}\otimes 1) \cdot B\big) .
	\end{split}
\end{align}
The proof of~\eqref{automorphisms_as_adj_of_dehn_twists} essentially repeats\footnote{The only difference with the elements defined in~\cite[Lem.\,9]{VS} is in normalization factor which we omit so that ${\hat v}(a)$ and ${\hat v}(b)$ are invertible.} the one of~\cite[Lem.\,9]{VS}, and so we omit it.
We will interpret the elements ${\hat v}(a)$ and ${\hat v}(b)$ as the ``quantum" Dehn twists operators along the $a-$ and $b-$cycles of the torus, correspondingly.

For the further analysis it will be important to note that the elements~\eqref{dehn_twist_quantum_eq1} actually belong to the gauge-invariant
subalgebra $\mathcal{A}$. Indeed, recall the standard result due to~\cite{[Drinfeld]}: let $H$ be a unimodular finite-dimensional Hopf algebra over $\mathbb{C}$ and $K\in H\tensor H$ such that $K\Delta(x) = \Delta(x)K$ for all $x\in H$, and let $\phi\colon H\to \mathbb{C}$ be a linear map such that
\be\label{eq:q-char-def}
\phi(xy) = \phi(S^2(y)x)
\ee
 then $(\phi\tensor id)(K)$ is in the centre of $H$.  Applying this to $K=(\ribbon^{-1}\otimes 1)\cdot M$ and $\phi = \mu$ (the equation~\eqref{eq:q-char-def} holds for the integral $\mu$, see~\cite{Ra-book}) we then get that
 $$
  (\mu\otimes id)\big((\ribbon^{-1}\otimes 1)\cdot M\big) \; \in \; Z(\cG)
 $$
 and therefore both the elements ${\hat v}(a)$ and ${\hat v}(b)$ are in the centres $Z\big(\cT^{(a)}\big)$ and $Z\big(\cT^{(b)}\big)$, respectively.
 Using the result in~\eqref{eq:Z-ab-in-A}, we conclude that  both ${\hat v}(a)$ and ${\hat v}(b)$ belong to $\cA$. We note however that these elements are not necessarily in the centre of $\cA$.

\medskip
Using the special elements ${\hat v}(a)$ and ${\hat v}(b)$ in $\cA$, we can make a statement, which in the following chapters will
be treated very concretely in the cases of a  simple  ``toy" model based on a finite cyclic
group, and then for the $\UBgl$ case at a root of unity. Using the quantum Dehn twist
operators, we can define the elements that correspond to the actions of the $SL(2,\mathbb{Z})$ group:
\begin{align} \label{handle_alg_S_transf}
	\SSS &:= {\hat v}(b) {\hat v}(a) {\hat v}(b) , \\ \label{handle_alg_T_transf}
	\TTT &:= \big({\hat v}(a)\big)^{-1} .
\end{align}
Recall that both ${\hat v}(a)$ and ${\hat v}(b)$  belong to $\cA$, therefore the elements $\SSS$ and $\TTT$ are in the gauge-invariant subalgebra $\cA$ too.

While $\SSS$ and $\TTT$ provide a modular group action on the elements of the handle
algebra~$\cT$ they do not necessarily  furnish a projective representation of $SL(2,\mathbb{Z})$ on its
representation space~$\mathfrak{R}$. However,  $\SSS$ and $\TTT$ can be realised as operators on
the  subspace $\RA$ of  gauge-invariant states introduced in~\eqref{eq:RA-def}, recall also our result in~\eqref{eq:RA-Inv}. This realisation follows from the above result  that $\SSS,\TTT\in\cA$, and therefore
$$
D(\SSS)\colon \; \RA \to \RA, \qquad D(\TTT)\colon \; \RA \to \RA.
$$
  Moreover, we claim that $\SSS$ and $\TTT$ operators provide a  projective
representation of $SL(2,\mathbb{Z})$, i.e. they satisfy
\begin{align}
	&\DA(\SSS^4) = id, \\
	&\DA\bigl((\SSS\TTT)^3\bigr) = \lambda \DA(\SSS^2) ,
\end{align}
where $\lambda\in\mathbb{C}^{\times}$. This statement is strictly speaking  conjectural,
it naturally generalises~\cite[Thm.\,28]{Alekseev:1995rn} to not necessarily semisimple algebras,
and our conjecture is supported by a non-trivial example we demonstrate in Section~\ref{sec:4}.

\subsection{Lyubashenko--Majid $SL(2,\mathbb{Z})$ action on the centre}\label{sec:SL2Z-centre}
We recall here another construction of a (projective) $SL(2,\bZ)$ representation associated with~$\cG$.
Having a  ribbon  factorisable (super) Hopf algebra $\cG$, one can construct an infinite series of mapping
class group representations on certain spaces of intertwining operators~\cite{Lyubashenko:1995}. In
particular for a torus without punctures, we have a (projective) representation of the group $SL(2,\bZ)$
on the centre of $\cG$, see the original reference~\cite{Lyubashenko:1994ma}. We will now review this
construction for the case of torus, mainly following the more recent exposition~\cite{FGR}.

The construction involves three main ingredients: integral, monodromy matrix, and a ribbon element.
Let $Z(\cG)$ denotes the centre of $\cG$. The  $S$- and $T$-transformations from the modular group
acting on $Z(\cG)$ are defined as
\begin{align}
	\SSSZ(z) &= (\mu\otimes id)\big\{(S(z)\otimes 1)\cdot M\big\}, \qquad z\in Z(\cG).\label{eq:S-transf} \\
	\TTTZ(z) &= \ribbon^{-1} \, z , \label{eq:T-transf}
\end{align}
with the ribbon element $\ribbon$ defined in \eqref{ribbon}. These two linear maps provide a projective
representation of $SL(2,\bZ)$:
\begin{align*}
	&\SSSZ^4=id, &&(\SSSZ\TTTZ)^3 = \lambda {\SSSZ}^2 ,
\end{align*}
with some non-zero number  $\lambda$. It is known that in the case of a modular Hopf algebra $\cG$, i.e.\
in the semisimple case, such a representation of $SL(2,\bZ)$ is equivalent to the Reshetikhin-Turaev
construction \cite{Reshetikhin:1990pr,Reshetikhin:1991tc}, where the $S$-transformation is provided by the closure (taking
the quantum trace) of the double braiding of a pair of irreducible representations. It was demonstrated
that the Reshetikhin-Turaev construction is equivalent to the handle algebra construction
in~\cite[Thm.\,29]{Alekseev:1995rn}.

\subsection{Conjecture on equivalence of two  $SL(2,\bZ)$ actions} \label{sec:conj}
So far, we have defined two  $SL(2,\bZ)$ actions, one
based on the handle algebra in Section~\ref{sec:SL2Z-handle} and the Lyubashenko-Majid one in Section~\ref{sec:SL2Z-centre}, and both are realized on the same vector space -- the centre of $\cG$.
Recall our result in~\eqref{eq:RA-Z}. As we just mentioned for the semisimple case, it is known that the two actions agree projectively.
Let us make the following conjecture.

\medskip
\textbf{Conjecture:}\; \textit{Let $\cG$ be a finite-dimensional ribbon factorisable (super) Hopf algebra over $\mathbb{C}$. The two $SL(2,\bZ)$  representations, one defined in~\eqref{handle_alg_S_transf}-\eqref{handle_alg_T_transf} on $\RA\cong Z(\cG)$ and the other in~\eqref{eq:S-transf}-\eqref{eq:T-transf}, are projectively isomorphic.}

\medskip
We will next demonstrate on two examples the two constructions of (projective) $SL(2,\bZ)$ actions, one
based on the handle algebra and the Lyubashenko-Majid one, and show explicitly that they are indeed equivalent.
We begin with a ``toy" model based on a  cyclic group.

\section{Toy model --- the cyclic group case}\label{sec:toy}
In this section, we demonstrate the construction described in the previous section in the simplest possible case --  the choice of the Hopf algebra $\cG$ given by (the group algebra)  of the finite cyclic group $\mathbb{Z}_p$ with $p\in\mathbb{N}$ elements. We will denote this algebra by $\toyalg$. This is a semisimple algebra, while a non-semisimple case is considered in the next section.

The  Hopf algebra $\toyalg:= \mathbb{C} \mathbb{Z}_p$ is generated by $k$ with the only relation $k^p=1$. It has the basis
 $\{k^{n}\}_{n=0}^{p-1}$ with the commutative multiplication $k^{n} k^{m} = k^{n+m}$,
and the group-like co-product $\Delta$, the co-unit $\epsilon$ and the antipode $S$ such that
\begin{align}
 &\Delta(k^{n}) = k^{n}\otimes k^{n} , && \epsilon(k^{n}) = 1, &S(k^{n}) = k^{-n}.
\end{align}

We will also use the notation  $q=e^{2\pi i/p}$.
For the algebra $\toyalg$, the universal $R$-matrix is then
\begin{align}
 R = \frac{1}{p} \sum_{n,m=0}^{p-1} q^{-nm} k^{n} \otimes k^{m} .
\end{align}
It is straightforward to check the $R$-matrix axioms.

The monodromy matrix~\eqref{universal} is then
\begin{align}
 M = \frac{1}{p} \sum_{n,m=0}^{p-1} q^{-2nm} k^{2n} \otimes k^{2m} .
\end{align}
It is however non-degenerate for odd values of $p$ only. Indeed, the  monodromy matrix can be  rewritten as $M=\sum_{n=0}^{p-1}  k^{2n} \tensor e_n$
where we introduced the idempotents
$$
e_n = \frac{1}{p} \sum_{m=0}^{p-1} q^{-2nm} k^{2m}.
$$
It is clear that $e_n$, for $0\leq n\leq p-1$, form a basis in $\toyalg$ for odd values of $p$, while for even $p$ we have $e_{n+p/2}=e_n$. Similarly, $\{k^{2n} \}_{n=0}^{p-1}$ is a basis in $\toyalg$ for odd $p$ only. Therefore, the monodromy matrix takes the form~\eqref{eq:M-nondeg} for odd values of $p$ only, and  $\toyalg$ is thus factorisable. By this reason, we will assume below that $p$ is odd.

 For $\toyalg$, the integral~\eqref{integral} takes the well-known form:
\begin{align}\label{integral_fcg}
 \mu(k^{n}) = \mathcal{N} \delta_{n,0},
\end{align}
where we use the normalisation~\eqref{normalisation}  that gives
\begin{align}\label{normalisation_toymodel}
 \mathcal{N} = \sqrt{p} .
\end{align}
Moreover, the co-integral \eqref{eq:coint-def} is given by  (using also the normalisation $\mu(\coint)=1$)
 \be
  \coint= \frac{1}{\sqrt{p} } \sum_{n=0}^{p-1} k^{n} . \label{cointegral_fcg}
 \ee

Using the definitions \eqref{balancing} and \eqref{ribbon} one can find that the balancing element
 $\pivot = 1$
and the expression for the ribbon element $\ribbon$ is given by
\begin{align}
 \ribbon^{\pm1}= 
 \frac{i^{\mp\omega(p)}}{\sqrt{p}} \sum_{n=0}^{p-1} q^{\pm n^2} k^{2 n},
\label{ribbon_fcg}
\end{align}
where we used the ``Gauss sum" identity
\begin{align}\label{q_sum_identity_fcg}
\sum_{m=0}^{p-1} q^{m^2} = i^{\omega(p)} \sqrt{p} ,
\end{align}
with
$$
\omega(p) = \left\{ \begin{array}{c} 1, \qquad p \in 4\mathbb{Z}+3, \\ 0, \qquad p \in 4\mathbb{Z}+5. \end{array} \right.
$$

\subsection{Representations of $\toyalg$}

The finite cyclic group has a very simple representation theory, given that it is a commutative and co-commutative algebra. It admits only 1-dimensional irreducible representations $\pi_n \colon \toyalg \to \mathbb{C}$ which are
\begin{align}
 \pi_n(k) = q^{n}\ , \qquad n=0,\ldots,p-1\ .
\end{align}
 On those representations, we have
\begin{align*}
 (\pi_n\otimes id)R &= k^{n}, \\
 (\pi_n\otimes id)M &= k^{2n}.
\end{align*}
Moreover, since the balancing element is trivial and the algebra is non-graded, the quantum trace~\eqref{eq:q-tr} is simply the ordinary trace on 1-dimensional representations
\begin{align*}
 \text{str}_q(\pi_n(k^m)) = q^{nm} .
\end{align*}

\subsection{The handle algebra of $\toyalg$}

In this section, we solve the exchange equations \eqref{handlealg_firstexchangerelation}-\eqref{handlealg_thirdexchangerelation} which define the handle algebra $\cT$ commutation relations. Since these first two of those relations are identical to the relations for the monodromy matrix of our toy model algebra, we use the following Ansatz for the universal elements $A$ and $B$
\begin{align}
 A &= \frac{1}{p} \sum_{n,m=0}^{p-1} q^{-2nm} k^{2n} \otimes \big(k^{(a)}\big)^{2m} , \\
 B &= \frac{1}{p} \sum_{n,m=0}^{p-1} q^{-2nm} k^{2n} \otimes \big(k^{(b)}\big)^{2m} ,
\end{align}
where the subalgebra spanned by $\bigl\{ \bigl(k^{(a)}\bigr)^{n} \bigr\}_{n=0}^{p-1}$ is the algebra isomorphic to $\toyalg$ associated to the $a$-cycle,
while $\bigl\{ \bigl(k^{(b)}\bigr)^{n} \bigr\}_{n=0}^{p-1}$ --- the one associated to the $b$-cycle.

Because the algebra $\toyalg$ is commutative and the $R$-matrix is symmetric, the third exchange relation \eqref{handlealg_thirdexchangerelation} simplifies to
\begin{align}
 A_{13} B_{23} &= (R_{12})^2 B_{23} A_{13} .
\end{align}
One can show easily that this equation is satisfied when one imposes the following commutation relation on the elements of the handle algebra
\begin{align*}
 \bigl(k^{(a)}\bigr)^n \bigl(k^{(b)}\bigr)^m = q^{\frac{n m}{2}} \bigl(k^{(b)}\bigr)^m \bigl(k^{(a)}\bigr)^n .
\end{align*}
Indeed, it is the commutation relations for the elements of the Heisenberg double of  $\toyalg$ (recall the discussion above Section~\ref{gauge}). We  have thus found all the defining relations in $\cT$.

\subsection{Fock module $\mathfrak{R}$}
In the following we explain the construction of the representation $D$ of the handle algebra from Section~\ref{section.2.4}. This representation has a cyclic vector, the vacuum defined by the trivial action of the $b$-cycle elements:
\begin{align}
 D\bigl((k^{(b)})^m\bigr)|0\rangle = |0\rangle \ ,
\end{align}
which of course agrees with~\eqref{eq:B-act-R}.
The representation space  $\mathfrak{R}$ is spanned by vectors $\{ |n\rangle \}_{n=0}^{p-1}$ which are defined via application of the elements of the $a$-cycle subalgebra to the vacuum:
\begin{align}
 |n\rangle := D\bigl((k^{(a)})^{2n}\bigr)|0\rangle,
\end{align}
which follows from the definition \eqref{vectors_of_fock_module} and the form of the universal element $A$. We note that actually all powers of $k^{(a)}$, odd and even, appear here -- it is due to the relation $(k^{(a)})^p=1$ and the condition that $p$ is odd.
 From here, using the commutation relations in $\cT$,  one can calculate the action on arbitrary vectors of $\mathfrak{R}$
\begin{align}
 D\bigl((k^{(b)})^n\bigr)|m\rangle = q^{-n m} |m\rangle .
\end{align}

\subsection{Gauge-invariance conditions}

Now, we want to investigate the gauge-invariance conditions, explained in Section \ref{gauge}, and find the gauge-invariant subalgebra $\cA$ of the handle algebra. Because of the commutativity of $\toyalg$, the equation \eqref{action_of_cG_on_cT} gives
\begin{equation}
 k(k^{(i)}) = k^{(i)} ,
\end{equation}
for $i=a,b$. From this follow the commutation relations for the elements of the smash product ${\bar \cT}$
\begin{align}\begin{aligned}\label{GactiononT_fcg}
 \iota(k^{n}) \bigl(k^{(a)}\bigr)^m &= \bigl(k^{(a)}\bigr)^m \iota(k^{n}) ,\\
 \iota(k^{n}) \bigl(k^{(b)}\bigr)^m &= \bigl(k^{(b)}\bigr)^m \iota(k^{n}) ,\end{aligned}
\end{align}
for all $n,m=0,\ldots,p-1$. Because all elements of $\cT$ commute with all elements of $\{ \iota(x)|x\in\cG\}$,
 the gauge-invariant subalgebra is in fact isomorphic to the handle algebra itself
\begin{align}
 \mathcal{A} = \mathcal{T} .
\end{align}

Finally, we extend the representation $D$ of the handle algebra $\cT$ to the smash product ${\bar \cT}$ by
\begin{align*}
 D(\iota(k^{n}))|m\rangle = |m\rangle .
\end{align*}
where we also  used the relations~\eqref{GactiononT_fcg}.

 \subsection{$SL(2,\mathbb{Z})$ action from the handle algebra}\label{section.3.6}
In this section, we construct the projective  $SL(2,\mathbb{Z})$ representation via operators on the representation space of the gauge-invariant subalgebra, which is the handle algebra in this case.
We obtain the matrix coefficients of the $\SSS$ and $\TTT$ transformations and we verify that those transformations indeed satisfy the relations of $SL(2,\mathbb{Z})$.

Using the integral \eqref{integral_fcg} and the ribbon element \eqref{ribbon_fcg}, we get the explicit formulae for the quantum Dehn twist operators defined by~\eqref{dehn_twist_quantum_eq1}
\be
 {\hat v}(a) = \frac{1}{\sqrt{p}} \sum_{n=0}^{p-1} q^{n^2} \big(k^{(a)}\big)^{-2n}, \qquad
 {\hat v}(b) = \frac{1}{\sqrt{p}} \sum_{n=0}^{p-1} q^{n^2} \big(k^{(b)}\big)^{-2n} .
\ee
One can directly check that their adjoint actions  via the automorphisms $\alpha$ and $\beta$ given by~\eqref{automorphisms_as_adj_of_dehn_twists} indeed satisfy the equations~\eqref{quantumaction}.
The two Dehn twists are represented on the representation space $\mathfrak{R}$ by
\begin{align}
& D({\hat v}(a)) | n \rangle = \frac{1}{\sqrt{p}} \sum_{m=0}^{p-1} q^{m^2} |n - m\rangle, \\
 \begin{split}
 & D({\hat v}(b)) | n \rangle = i^{\omega(p)} q^{-n^2} |n \rangle,
 \end{split}
\end{align}
where we used the identity \eqref{q_sum_identity_fcg}. Using this representation, the $\SSS$- and $\TTT$-matrices, defined by \eqref{handle_alg_S_transf} and \eqref{handle_alg_T_transf} respectively, are realised as
\begin{align}
 &D(\SSS) | m \rangle = \frac{(-1)^{\omega(p)}}{\sqrt{p}} \sum_{n=0}^{p-1} q^{-2nm} |n \rangle ,\\
 &D(\TTT) | m \rangle = \frac{1}{\sqrt{p}} \sum_{n=0}^{p-1} q^{-n^2} |m + n\rangle .
 \end{align}
From this explicit action, one can easily calculate that
\begin{align}
& D(\SSS^2) | m \rangle = |-m\rangle,
 &&D(\SSS^4) | m \rangle = |m\rangle ,
\end{align}
and by iteratively applying the above expressions  we get
\begin{align}
D\big((\SSS\TTT)^3\big) | m \rangle & = |-m\rangle.
\end{align}
By comparing the expressions, we see that the $SL(2,\mathbb{Z})$ relations are indeed satisfied:
\begin{align}
&(\SSS\TTT)^3 =  \SSS^2,
 &&\SSS^4 = id.
\end{align}

 \subsection{Lyubashenko--Majid $SL(2,\mathbb{Z})$ action on the centre}\label{section.3.7}
As the algebra $\toyalg$ is commutative, the centre $Z(\toyalg)$ is $\toyalg$. However, we note that the canonical construction of central elements via the map defined in~\eqref{eq:z-pi} gives
 \begin{align*}
 a_n \equiv z_{\pi_n} =  (\pi_n\otimes id)(M) = k^{2n}.
 \end{align*}

 By using equation \eqref{eq:S-transf} one can find the $\SSS$ transformation on the central elements
 \begin{align}
 \SSSZ(a_n) = \frac{1}{\sqrt{p}} \sum_{m=0}^{p-1} q^{-2nm} a_m,
 \end{align}
 and therefore
 \begin{align*}
 &\SSSZ^2(a_n) = a_{-n} ,
 &&\SSSZ^4(a_n) = a_n .
 \end{align*}
 In addition we have $\TTTZ$ transformation as it was defined in the equation \eqref{eq:T-transf}
\begin{align}
\TTTZ(a_n) = \frac{i^{\omega(p)}}{\sqrt{p}} \sum_{m=0}^{p-1} q^{-(m-n)^2} a_{m} .
\end{align}
Therefore
\begin{align*}
& (\SSSZ\TTTZ)(a_n) = \frac{1}{\sqrt{p}} \sum_{m=0}^{p-1} q^{m^2-2nm} a_{m} ,
&&~~~~~~~~~(\SSSZ\TTTZ)^3(a_n) = i^{\omega(p)} a_{-n} ,
\end{align*}
where we used again  the identity \eqref{q_sum_identity_fcg}.
Therefore, we  get  the relation
\begin{align}
(\SSSZ\TTTZ)^3 = \lambda {\SSSZ}^2 ,
\end{align}
for  $\lambda = i^{\omega(p)}$, i.e.\ we have indeed a projective representation of $SL(2,\bZ)$.

\subsection{Equivalence of two actions} \label{Equivalence_fcg}
In this section, we show that the two $SL(2,\bZ)$ actions presented in Sections~\ref{section.3.6} and~\ref{section.3.7} agree projectively. In order to do that, we first establish that the centre  $Z(\toyalg)=\toyalg$  and the representation space $\mathfrak{R}$ of the gauge-invariant algebra $\cA$  for $\cG=\toyalg$ are isomorphic as vector spaces, and this of course agrees with our general result established in~~\eqref{eq:RA-Z}.

Explicitly, we have that
\begin{align}
Z(\toyalg) \ni a_n \stackrel{\simeq}{\longmapsto} |n\rangle \in \mathfrak{R} .
\end{align}
Moreover, if we take into account this isomorphism, we can compare the coefficients of the relevant actions in the two cases. In order to do that, let us define the coefficients of the $\SSS$- and $\TTT$-actions as
\begin{align*}
D(\SSS)|m\rangle &= \sum_{n=0}^{p-1} (\SSS_\mathcal{T})^n_m |n\rangle, && D(\TTT)|m\rangle = \sum_{n=0}^{p-1} (\TTT_\mathcal{T})^n_m |n\rangle,
\end{align*}
for the handle algebra and
\begin{align*}
\SSSZ(a_m) &= \sum_{n=0}^{p-1} (\SSSZ)^n_m a_n, && \TTTZ(a_m) = \sum_{n=0}^{p-1} (\TTTZ)^n_m a_n,
\end{align*}
for the centre of $\toyalg$ in the LM picture. It is easy to read-off that those coefficients are
\begin{align*}
(\SSS_\mathcal{T})^n_m &= (-1)^{\omega(p)} (\SSSZ)^n_m = (-1)^{\omega(p)} \frac{1}{\sqrt{p}} q^{-2nm}, \\
(\TTT_\mathcal{T})^n_m &= i^{-\omega(p)} (\TTTZ)^n_m = \frac{1}{\sqrt{p}} q^{-(n-m)^2} .
\end{align*}
We see therefore that those two actions agree up to multiplicative constants, i.e. they agree projectively, as it was claimed.

\section{$\UBgl$ algebra and its representation}\label{sec:4}

\def\UBglsimple{\cG}

In this section, we introduce a restricted version of the quantum enveloping Hopf algebra $\UBgl$ with $q$ being the primitive $p$th root of unity, where $p$ is an odd integer. To simplify notation, it will be understood that within this section $\cG=\UBgl$.

We begin in Section~\ref{sec:4-relations} with recalling the Hopf algebra structure on $\cG$ and compute its (co)integrals. Then in Section~\ref{sec:R-v}, we introduce the ribbon structure:  we give the universal $R$-matrix, and calculate the corresponding monodromy matrix and the ribbon element.  In Section~\ref{rep}, we also review known facts about the representation theory of this algebra. Then we construct the corresponding handle algebra $\cT$ in Section~\ref{sec:handle_gl11} in terms of generators and relations, its gauge-invariant subalgebra $\cA$ is studied in Section~\ref{sec:4-A}, and  its Fock module $\RA$ is described in Section~\ref{sec:4-Fock}.
The $SL(2,\mathbb{Z})$ action from the handle algebra approach is analysed in Section~\ref{sec:handle_mcg_gl11} where we also establish an equivalence with the modular action in~\cite{Mikhaylov:2015qik}. Finally,  in Section~\ref{sec:4-comparison} we compare this action to the Lyubashenko--Majid action of $SL(2,\mathbb{Z})$ analysed in Section~\ref{sec:LM_mcg_gl11}, confirming the conjecture formulated in Section~\ref{sec:conj}.

\subsection{Definition and (co)integrals}\label{sec:4-relations}

The restricted quantum group for $gl(1|1)$ that will be denoted by  $\cG=\UBgl$ is a super Hopf algebra  generated by $\ka$, $\kb$ and $e_+$, $e_-$ with the defining relations
\begin{align}\begin{aligned}
 &\ka^p=\kb^p=1, \\ & \ka e_\pm = e_\pm \ka, \\ & \kb e_\pm = q^{\pm1} e_\pm \kb,
\end{aligned} && \begin{aligned}
 & \ka \kb = \kb \ka, \\ & \{e_\pm,e_\pm\} = 0, \\ & \{e_+,e_- \} =\frac{\ka-\ka^{-1}}{q-q^{-1}}, \end{aligned}
\end{align}
where the  parameter is $q=e^{2\pi i/p}$ and we assume $p$ is a positive odd integer. We note that the generator $\ka$ is central. It is a finite-dimensional algebra with the basis
\be\label{eq:seq4-basis}
\{ \ka^n \kb^m e_+^r e_-^s \ | \ 0\leq n,m\leq p-1 \;,\; 0\leq r,s\leq 1\}\ .
\ee
 The co-product has the form
\begin{align}\begin{aligned}
 \Delta(\ka) &= \ka\otimes\ka, \\ \Delta(\kb) &= \kb\otimes\kb,
\end{aligned} && \begin{aligned}
 &\Delta(e_+)= e_+\otimes1 + \ka^{-1}\otimes e_+, \\ &\Delta(e_-)= e_-\otimes \ka + 1\otimes e_-, \end{aligned}
\end{align}
the co-unit is
\begin{align}
 \epsilon(\ka) = \epsilon(\kb) &= 1, && \epsilon(e_+) = \epsilon(e_-) = 0,
\end{align}
and the antipode is
\begin{align}\begin{aligned} S(\ka) &= \ka^{-1}, \\ S(\kb) &= \kb^{-1}, \end{aligned} && \begin{aligned} &S(e_+) = -\ka e_+, \\ &S(e_-)= - e_- \ka^{-1} . \end{aligned}
\end{align}

The right integral as it was defined in \eqref{integral} evaluated on the basis~\eqref{eq:seq4-basis} of $\UBglsimple$ has the following form
\begin{align}
 \mu( \ka^{n} \kb^{m} e_+^r e_-^s ) = \mathcal{N} \delta_{n,-1} \delta_{m,0} \delta_{r,1} \delta_{s,1} ,
\end{align}
or alternatively we can write it as
\begin{equation}\label{eq:sec4-int}
 \mu = \mathcal{N} ( \ka^{-1} e_+ e_- )^* ,
\end{equation}
where the normalisation $\mathcal{N}$ will be fixed later.

It can be easily checked that the co-integral~\eqref{eq:coint-def} is given by
 \be
  \coint= \frac{1}{\mathcal{N} } \sum_{n,m=0}^{p-1} \ka^{n} \kb^m e_+ e_- , \label{cointegral_gl11}
 \ee
 where as usual we normalise it by $\mu(\coint)=1$. It is a two-sided co-integral.

\subsection{$R$-matrix and ribbon element}\label{sec:R-v}
The super Hopf algebra $\cG$ is quasi-triangular with the universal $R$-matrix
\begin{align}
 R = \frac{1}{p^2} \left( 1\otimes 1 - (q-q^{-1}) e_+\otimes e_- \right) \sum_{n,m=0}^{p-1} \sum_{s,t=0}^{p-1} q^{nt+ms} \ka^{n}\kb^{m} \otimes \ka^{-s}\kb^{-t} .
\end{align}
This form of $R$-matrix was motivated by the construction \cite{Khoroshkin:1991bka}  in the case of  generic values of $q$.
In addition, it will be useful (for the handle algebra relations) to spell explicitly the inverse of the $R$-matrix
\be
 R^{-1}
 = \frac{1}{p^2} \left( 1\otimes 1 + (q-q^{-1}) \ka e_+\otimes \ka^{-1} e_- \right) \sum_{n,m=0}^{p-1} \sum_{s,t=0}^{p-1} q^{-nt-ms} \ka^{n}\kb^{m} \otimes \ka^{-s}\kb^{-t} .
\ee
And we also need the monodromy matrix
\begin{equation}\label{eq:sec4-M}
\begin{split}
 M =\,&  \big( (q-q^{-1})^{-1}1\otimes 1 + e_-\otimes e_+ - \ka^{-1} e_+\otimes \ka e_- + (q-q^{-1}) \ka^{-1} e_-e_+\otimes \ka e_+e_- )  \\
 &\times (q-q^{-1})\frac{1}{p^2} \sum_{n,m,s,t=0}^{p-1} q^{-2nt-2ms} \ka^{2n}\kb^{2m} \otimes \ka^{2s}\kb^{2t}. \end{split}
\end{equation}

Introducing the idempotents
$$
e_{n,m} =\frac{1}{p^2} \sum_{s,t=0}^{p-1} q^{-2nt-2ms} \ka^{2s} \kb^{2t},
$$
the second line in~\eqref{eq:sec4-M} can be written as $(q-q^{-1}) \sum_{n,m=0}^{p-1}  \ka^{2n}\kb^{2m} \otimes e_{n,m}$. Then similarly to analysis of $M$ in Section~\ref{sec:toy},
we conclude that $M$ takes the form~\eqref{eq:M-nondeg}, and it is thus non-degenerate. We note that this is true  for odd $p$ only, because only then $\{\ka^{2n} \kb^{2m} e_+^r e_-^s \}_{n,m,r,s=0}^{p-1,p-1,1,1}$ and $\{e_{n,m} e_+^r e_-^s \}_{n,m,r,s=0}^{p-1,p-1,1,1}$ are  bases of $\UBglsimple$. (This is why we assumed above that $p$ is odd.) Therefore, $\UBglsimple$ is a factorisable super Hopf algebra.

With the monodromy matrix~\eqref{eq:sec4-M} and according to the equation \eqref{normalisation} we can fix the normalisation for the integral in~\eqref{eq:sec4-int} as
\begin{align}
 \mathcal{N} = \frac{i p}{q-q^{-1}} .
\end{align}

Using the right integral $\mu$ from~\eqref{eq:sec4-int}, we find the co-modulus $\comod \in\mathcal{G}$ \eqref{left_integral} to be
\begin{equation}
 \comod = \ka^{-2} ,
\end{equation}
which admits a group-like square root, and therefore the balancing element $\pivot$ is just
\begin{equation}
\pivot = \ka^{-1} .
\end{equation}
We note that the element $\pivot$ is central and it satisfies $S^2(x)= \pivot x {\pivot}^{-1}$ for all $x\in\UBglsimple$, and this is consistent with the fact that $S^2 = id$ in this case.

Then, using the expression for the ribbon element $\ribbon$ from \eqref{ribbon} we find its explicit form (after making appropriate re-summation)
\begin{equation} \begin{split}
\ribbon^{\pm1}&= \frac{1}{p} \ka^{\pm1} (1\mp(q-q^{-1}) \ka^{\mp1} e_-e_+) \sum_{n,m=0}^{p-1} q^{\pm2nm} \ka^{2n} \kb^{2m} . \end{split}
\end{equation}
One can of course directly check the ribbon axioms~\eqref{eq:rib-ax}. We therefore conclude that $\cG$ is a ribbon factorisable super Hopf algebra.

\subsection{Representations of $\UBgl$}\label{rep}
Here, we briefly review representation theory of $\UBglsimple$, which has been studied e.g. in  \cite{Rozansky:1992rx,Schmidke:1990ve,Corrigan:1989ev,Chaichian:1990un}.

The important class of representations consists of  1-dimensional atypical representations, 2-dimensional typical representations and 4-dimensional indecomposable projective representations\footnote{These are actual linear representations,  also often  called ``projective modules",  and should not be confused with  {\sl projective} representations from group theory which are linear up to a multiplicative constant.} that we describe below in a basis.
 The major difference from the previous section when we considered the algebra $\toyalg$ based on the finite cyclic group  is that  $\UBgl$ is not semisimple --- as we recall below, there are 4-dimensional  projective representations which are reducible but indecomposable.

We start describing the so-called atypical representations $\pi_n \colon\;  \UBglsimple \to \mathbb{C}^{1|0}$, for $n=0,\ldots,p-1$,
\begin{align}\begin{aligned}
 &\pi_n(\ka) = 1, \\ & \pi_n(\kb) = q^n,
\end{aligned} && \begin{aligned}
 \pi_n(e_+) = 0, \\ \pi_n(e_-) = 0,
\end{aligned}\end{align}
All the atypical  representations are one-dimensional and clearly  irreducible. The co-unit $\epsilon$ corresponds to the atypical representation $\pi_0$. Moreover, we have a series of the so-called typical representations $\pi_{e,n} \colon\;  \UBglsimple \to \End(\mathbb{C}^{1|1})$ with $e,n=0,\ldots,p-1$
\begin{align}\begin{aligned} &\pi_{e,n}(\ka) = q^e \left( \begin{array}{cc} 1 & 0 \\ 0 & 1 \end{array} \right), \\ & \pi_{e,n}(\kb) = q^n \left( \begin{array}{cc} q^{-1} & 0 \\ 0 & 1 \end{array} \right), \end{aligned} && \begin{aligned} &  \pi_{e,n}(e_+) = \left( \begin{array}{cc} 0 & 0 \\ {[e]_q} & 0 \end{array} \right), \\ & \pi_{e,n}(e_-) = \left( \begin{array}{cc} 0 & 1 \\ 0 & 0 \end{array} \right), \end{aligned}
\end{align}
where $[x]_q = \frac{q^x - q^{-x}}{q-q^{-1}}$ is the $q$-number. The typical representations $\pi_{e,n}$ are two-dimensional (with 1 even and 1 odd degrees), and when $e\neq0$ they are irreducible. When $e=0$, the representation $\pi_{0,n}$ is not irreducible anymore, but it is still indecomposable: it is built up from two atypical irreducible representations $\pi_n$ and $\pi_{n-1}$ connected by the action of $e_-$. The corresponding subquotient structure  can be written diagrammatically  as
\begin{align}
 \pi_n \to \pi_{n-1},
\end{align}
where the arrow points to a submodule and it corresponds to a ``non-invertible" action of the algebra.

Besides the atypical and typical representations, one has as well the projective representations $\pi_{\mathcal{P}_N} \colon\;  \UBglsimple \to Hom(\mathbb{C}^{2|2})$, which are defined by the matrix realisations on 4-dimensional vector space with 2 even
and 2 odd
degrees as follows
\begin{align}\begin{aligned} & \pi_{\mathcal{P}_N}(\ka) = \left( \begin{array}{cccc} 1 & 0 & 0 & 0 \\ 0 & 1 & 0 & 0 \\ 0 & 0 & 1 & 0 \\ 0 & 0 & 0 & 1 \end{array} \right) , \\ &\pi_{\mathcal{P}_N}(\kb) = q^N \left( \begin{array}{cccc} q^{-1} & 0 & 0 & 0 \\ 0 & 1 & 0 & 0 \\ 0 & 0 & 1 & 0 \\ 0 & 0 & 0 & q \end{array} \right), \end{aligned} && \begin{aligned} & \pi_{\mathcal{P}_N}(e_+) = \left( \begin{array}{cccc} 0 & 0 & 0 & 0 \\ -q^{-1} & 0 & 0 & 0 \\ 1 & 0 & 0 & 0 \\ 0 & 1 & q^{-1} & 0 \end{array} \right) ,\\ & \pi_{\mathcal{P}_N}(e_-) = \left( \begin{array}{cccc} 0 & 1 & q^{-1} & 0 \\ 0 & 0 & 0 & q^{-1} \\ 0 & 0 & 0 & -1 \\ 0 & 0 & 0 & 0 \end{array} \right), \end{aligned}
 \end{align}
where vectors $(1 \, 0 \, 0 \, 0)^t$ and $(0 \, 0 \, 0 \, 1)^t$ are odd and $(0 \, 1 \, 0 \, 0)^t$ and $(0 \, 0 \, 1 \, 0)^t$ are even, and where $^t$ denotes the transposition.

The representations $\pi_{\mathcal{P}_N}$ are reducible but indecomposable, and come from the tensor product of two typical representations $\pi_{-e,N}\otimes \pi_{e,1}$.
 They are built up from 4 atypical representations $\pi_{N+1}, \pi_N, \pi_N, \pi_{N-1}$ which constitute the module according to the subquotient diagram
\begin{equation}\label{eq:PN-diag}
\begin{split}
\xymatrix{
&&\\
&\pi_{\mathcal{P}_N} \quad = \; &\\
&&}
 \xymatrix{ & \pi_N \ar[dr] \ar[dl] & \\ \pi_{N+1} \ar[dr] & & \pi_{N-1} \ar[dl] \\ & \pi_N \ar[uu]_\sigma & }\end{split}
\end{equation}
where the arrows are meant to be actions of $e_\pm$ (and here the map $\sigma$ should be ignored for a moment, it will be explained later). Explicitly, the following vectors of the 4-dimensional module constitute the modules of the atypical representations in the diagram
\begin{align*}
 &\left( \begin{array}{c} 1 \\ 0 \\ 0\\ 0 \end{array} \right) \in \pi_{N-1} ,
  &&\left( \begin{array}{c} 0 \\ -q^{-1} \\ 1 \\ 0 \end{array} \right) \in \text{bottom } \pi_N , %\\
 &\left( \begin{array}{c} 0 \\ 0 \\ 0\\ 1 \end{array} \right) \in \pi_{N+1} ,
 &&\left( \begin{array}{c} 0 \\ 1 \\ q \\ 0 \end{array} \right) \in \text{top } \pi_N .
 \end{align*}

 It is worthwhile to note that the Casimir element of $\UBglsimple$ evaluated on the projective representation $\pi_{\mathcal{P}_N}$ maps the top atypical representation to the bottom one, and it is zero otherwise. It is not realised by an invertible matrix.

For the purposes of the next section, we want to find a matrix $\sigma$ that maps the bottom atypical sub-representation to the top one and it is zero otherwise, i.e.\ $\sigma$ satisfies the following relations
\begin{align*}
 &\sigma \left( \begin{array}{c} 0 \\ -q^{-1} \\ 1 \\0 \end{array} \right) = \left( \begin{array}{c} 0 \\ 1 \\ q \\ 0 \end{array} \right) ,
 &&\sigma \left( \begin{array}{c} 0 \\ 1 \\ q \\ 0 \end{array} \right) = \left( \begin{array}{c} 0 \\ 0 \\ 0 \\0 \end{array} \right) ,
 &\sigma \left( \begin{array}{c} 1 \\ 0 \\ 0 \\0 \end{array} \right) = \left( \begin{array}{c} 0 \\ 0 \\ 0 \\ 0 \end{array} \right) ,
 &&\sigma \left( \begin{array}{c} 0 \\ 0 \\ 0 \\ 1 \end{array} \right) = \left( \begin{array}{c} 0 \\ 0 \\ 0 \\0 \end{array} \right) ,
\end{align*}
which determines $\sigma$ up to a multiplicative constant. It is realised as a matrix
\begin{align}\label{eq:sigma-map}
 \sigma &= \frac{q}{2} \left( \begin{array}{cccc} 0 & 0 & 0 & 0 \\ 0 & -1 & q^{-1} & 0 \\ 0 & -q & 1 & 0 \\ 0 & 0 & 0 & 0 \end{array} \right) .
\end{align}

Finally, let us recall the definition of the quantum supertrace. The ordinary supertrace of an $n\times n$ matrix $\mathbf{X}$ with the coefficients $[ \mathbf{X} ]^i_j = X^i_j$ is defined as
\begin{align}
 \text{str}( \mathbf{X} ) = \sum_{i=1}^{n} (-1)^{|i|} X^i_i,
\end{align}
where $|i|$ denotes the grading of the diagonal element $X^i_i$. As applied to two-dimensional matrices on $\mathbb{C}^{1|1}$, the supertrace is explicitly
\begin{align*}
 \text{str} \left( \begin{array}{cc} x^1_1 & x^1_2 \\ x^2_1 & x^2_2 \end{array} \right) = x^1_1-x^2_2,
\end{align*}
and for four-dimensional matrices on $\mathbb{C}^{2|2}$
\begin{align*}
 \text{str} \left( \begin{array}{cccc} x^1_1 & x^1_2 & x^1_3 & x^1_4 \\ x^2_1 & x^2_2 & x^2_3 & x^2_4 \\ x^3_1 & x^3_2 & x^3_3 & x^3_4 \\ x^4_1 & x^4_2 & x^4_3 & x^4_4 \end{array} \right)  = x^1_1-x^2_2-x^3_3+x^4_4 .
\end{align*}
Then, the quantum supertrace is defined by inserting the inverse balancing element $\pivot^{-1}$:
\begin{align}
 \text{str}_q(\pi(X)) = \text{str}(\pi(\pivot^{-1} X)) ,
\end{align}
where $\pi$ is, for our purposes, a representation of $\UBglsimple$, i.e. $\pi=\pi_n,\pi_{e,n},\pi_{\mathcal{P}_N}$.

\subsection{The centre of $\UBglsimple$}\label{sec:4-centre}
In this section, we construct a basis in the centre $Z(\UBglsimple)$ of  $\UBglsimple$ using the description of projective representations in the previous section.

First, we recall that central elements can be constructed using the so-called Drinfeld map
$$
\phi \mapsto (\phi\tensor id) (M)
$$
 for $\phi$ satisfying~\eqref{eq:q-char-def}. Examples of such $\phi$ are the quantum traces $\str_q$ over representations of $\cG$, recall~\eqref{eq:q-tr}. Using this construction for the typical and atypical representations,  we obtain the central elements
\begin{align}\label{basis_of_the_centre_of_gl11_eq1}
 c_{e,n} &= (\str_q\otimes id)\big\{(\pi_{e,n}\otimes id) M\big\} , \qquad  1\leq e\leq p-1 , \; 0\leq n\leq p-1.\\
 a_{n} &= (\str_q\otimes id)\big\{(\pi_{n}\otimes id) M\big\}, \qquad \; 0\leq n\leq p-1,
\end{align}
$p^2$ elements in total.
Their span however gives only a proper subalgebra in $Z(\UBglsimple)$. To construct  the missing central elements, we follow an approach in~\cite{Gainutdinov:2007tc} that uses the so-called   pseudo-traces. For this, one has to consider a direct sum of all projective indecomposables from the same (categorical) block which is not semisimple. We recall that each block corresponds to a two-sided ideal in the algebra and vice versa, and in our case we have just one such non-semisimple block (in contrast to~\cite{Gainutdinov:2007tc} where one had to consider $p-1$ of them). We thus have only one missing central element and it is given by
 \be
 b = (\str_q\otimes id)\big\{(\pi_\sigma\otimes id) M\big\},\label{basis_of_the_centre_of_gl11_eq3}
\ee
where we introduced a special map $\pi_\sigma\colon \cG \to \mathrm{End}(\oplus_{N}\mathcal{P}_N)$ as
\be
\pi_\sigma(-):= \bigoplus_{N=0}^{p-1}\sigma\circ\pi_{\mathcal{P}_N} (-)
\ee
and the linear map $\sigma$ is defined in~\eqref{eq:PN-diag}-\eqref{eq:sigma-map}. (The quantum trace $\str_q$ precomposed with such a map is what we call pseudo-traces, see more details in~\cite{Gainutdinov:2007tc}.)

Explicitly, the central elements introduced above are given by the following expressions
\begin{align}
 c_{e,n} &= q^e (q-q^{-1}) \left\{ (1-q^{-2e}) e_+ e_- - \frac{\ka-\ka^{-1}}{q-q^{-1}} \right\} \ka^{2n-1} \kb^{2e} , \label{eq:central-1}\\
 a_{n} &= \ka^{2 n} , \\
 b &= -2(q-q^{-1})^2 \sum_{t=0}^{p-1} \ka^{2t} e_+ e_- .\label{eq:central-3}
\end{align}
We first note that these are linearly independent elements in the centre of $\UBglsimple$.
To show that any central element is a linear combination of these ones, we first calculate  dimension of the centre by analysing bimodule endomorphisms of the regular representation along the lines in~\cite{Feigin:2005zx}, and conclude that the dimension
\be
\dim\big(Z(\cG)\big) = p^2+1
\ee
agrees with the number of the central elements given above.

\subsection{The handle algebra of $\UBgl$}\label{sec:handle_gl11}

In this section, we describe the handle algebra of $\UBglsimple$ in terms of generators and relations.

We start by stating the form of the universal elements $A$ and $B$ corresponding to the $a$- and $b$-cycles, which solve the exchange equations \eqref{handlealg_firstexchangerelation}-\eqref{handlealg_secondexchangerelation}:
\begin{align*}\begin{split}
A =&\,  \left[ (q-q^{-1})^{-1} + e_-\otimes e_+^{(a)} - \ka^{-1} e_+\otimes \ka^{(a)}e_-^{(a)} + (q-q^{-1}) \ka^{-1} e_-e_+\otimes \ka^{(a)} e_+^{(a)}e_-^{(a)} \right] %\times
\\
&\times \frac{1}{p^2} (q-q^{-1}) \sum_{n,m,s,t=0}^{p-1} q^{-2nt-2ms} \ka^{2n} \kb^{2m} \otimes \big(\ka^{(a)}\big)^{2s} \big(\kb^{(a)}\big)^{2t}, \end{split} \\ \begin{split}
B =\,&\left[ (q-q^{-1})^{-1} + e_-\otimes e_+^{(b)} - \ka^{-1} e_+\otimes \ka^{(b)}e_-^{(b)} + (q-q^{-1}) \ka^{-1} e_-e_+\otimes \ka^{(b)} e_+^{(b)}e_-^{(b)} \right]% \times
 \\
&\times  \frac{1}{p^2} (q-q^{-1})  \sum_{n,m,s,t=0}^{p-1} q^{-2nt-2ms} \ka^{2n} \kb^{2m} \otimes \big(\ka^{(b)}\big)^{2s} \big(\kb^{(b)}\big)^{2t}. \end{split}
\end{align*}
where the elements
$$
\big\{ \big(\ka^{(a)}\big)^{n}\big(\kb^{(a)}\big)^{m}\big(e_+^{(a)}\big)^{r}\big(e_-^{(a)}\big)^{s} \big\}_{n,m,r,s=0}^{p-1,p-1,1,1}
$$
 span the subalgebra $\cT^{(a)}$  isomorphic to $\mathcal{G}$ associated to the $a$-cycle, and
 $$
 \big\{ \big(\ka^{(b)}\big)^{n}\big(\kb^{(b)}\big)^{m}\big(e_+^{(b)}\big)^{r}\big(e_-^{(b)}\big)^{s} \big\}_{n,m,r,s=0}^{p-1,p-1,1,1}
 $$
  --- the one associated to the $b$-cycle.

One can show that the third exchange relation \eqref{handlealg_thirdexchangerelation} implies the following ``mixed" commutation relations
\begin{align}\begin{aligned}
&\big(\kb^{(a)}\big)^{2a} \big(\kb^{(b)}\big)^{2b} = \big(\kb^{(b)}\big)^{2b} \big(\kb^{(a)}\big)^{2a} , \\ &\big(\kb^{(a)}\big)^{2a} \big(\ka^{(b)}\big)^{2b} = q^{2ab} \big(\ka^{(b)}\big)^{2b} \big(\kb^{(a)}\big)^{2a} , \\ &\big(\kb^{(a)}\big)^{2a} e_-^{(b)} = q^{-a} e_-^{(b)} \big(\kb^{(a)}\big)^{2a} ,\\ &\big(\kb^{(b)}\big)^{2a} e_+^{(a)} = q^{2a} e_+^{(a)} \big(\kb^{(b)}\big)^{2a} , \\ & \\ &\big[\big(\ka^{(a)}\big)^a, e_+^{(b)}\big] = 0, \\ &\big[\big(\ka^{(a)}\big)^a, e_-^{(b)}\big] = 0, 
\end{aligned} && \begin{aligned}
&\big(\ka^{(a)}\big)^{2a} \big(\ka^{(b)}\big)^{2b} = \big(\ka^{(b)}\big)^{2b} \big(\ka^{(a)}\big)^{2a} , \\ &\big(\ka^{(a)}\big)^{2a} (\kb^{(b)})^{2b} = q^{2ab} (\kb^{(b)})^{2b} (\ka^{(a)})^{2a} , \\ &\big[\big(\kb^{(a)}\big)^{2a}, e_+^{(b)}\big] = q^{a} [a]_q(q-q^{-1}) e_+^{(a)} \big(\kb^{(a)}\big)^{2a}, \\ &\big(\kb^{(b)}\big)^{2a} e_-^{(a)} = q^{-a} e_-^{(a)} \big(\kb^{(b)}\big)^{2a} + \\ &- q^{-2a}[a]_q (q-q^{-1}) (\ka^{(a)}) \big(\ka^{(b)}\big)^{-1} e_-^{(b)} (\kb^{(b)})^{2a} , \\ &\big[\big(\ka^{(b)}\big)^a, e_+^{(a)}\big] = 0, \\ &\big[\big(\ka^{(b)}\big)^a, e_-^{(a)}\big] = 0,
\end{aligned}
\end{align}
and the following anti-commutation relations
\begin{align}\begin{aligned}
& \{e_+^{(a)}, e_+^{(b)} \} = 0,~~~~~~~~~&&\{e_+^{(a)}, e_-^{(b)}\} = \ka^{(b)} (q-q^{-1})^{-1},\\
&\{e_-^{(a)}, e_-^{(b)}\} = 0,
~~~~~~~~~&& \{e_-^{(a)}, e_+^{(b)} \} = \left( \ka^{(a)} - \big(\ka^{(a)}\big)^{-1} - \ka^{(a)}\big (\ka^{(b)}\big)^{-2}\right) (q-q^{-1})^{-1}.
\end{aligned}
\end{align}
We claim that the above relations, together with the  $\UBgl$ relations for the generators of the subalgebras  $\cT^{(a)}$ and $\cT^{(b)}$ due to the isomorphisms noted above, constitute the complete set of defining relations for $\cT$.

\subsection{The  action of $\cG$ on $\cT$}
Now, we want to investigate the left action of $\cG$ on $\cT$, with the end-goal of constructing the smash product ${\bar \cT}$. Using~\eqref{action_of_cG_on_cT} we obtain
\begin{align*}
 \begin{aligned}
  \ka (\ka^{(i)} ) &= \ka^{(i)}, \\ \kb (\ka^{(i)} ) &= \ka^{(i)} , \\ e_+(\ka^{(i)}) &= 0 , \\ e_-(\ka^{(i)} ) &= 0,
 \end{aligned} &&
 \begin{aligned}
  \ka (\kb^{(i)} ) &= \kb^{(i)}, \\ \kb (\kb^{(i)} ) &= \kb^{(i)} , \\ e_+(\kb^{(i)}) &= \big[e_+^{(i)}, \kb^{(i)}\big] , \\
  e_-(\kb^{(i)}) &= \big[e_-^{(i)}, \kb^{(i)}\big] \big(\ka^{(i)}\big)^{-1},
 \end{aligned} &&
 \begin{aligned}
  \ka (e_+^{(i)}) &= e_+^{(i)}, \\ \kb(e_+^{(i)}) &= q e_+^{(i)} , \\ e_+(e_+^{(i)}) &= 0 , \\ e_-(e_+^{(i)}) &= \ffrac{1-\big(\ka^{(i)}\big)^{-2}}{q-q^{-1}},
 \end{aligned} &&
 \begin{aligned}
  \ka(e_-^{(i)}) &= e_-^{(i)}, \\ \kb(e_-^{(i)}) &= q^{-1} e_-^{(i)} , \\ e_+(e_-^{(i)}) &= \ffrac{\ka^{(i)}-\big(\ka^{(i)}\big)^{-1}}{q-q^{-1}} , \\ e_-(e_-^{(i)}) &= 0,
 \end{aligned}
\end{align*}
for $i=a,b$, and with the obvious choice of algebra isomorphisms $\kappa^{(i)}\colon\; \mathcal{G}\to \cT^{(i)}$
\begin{align}\label{embeddings_of_G_in_T}
 \kappa^{(i)}(\ka) = \ka^{(i)}, \quad \kappa^{(i)}(\kb) = \kb^{(i)}, \quad \kappa^{(i)}(e_\pm) = e_\pm^{(i)} .
\end{align}
This leads to the following (anti-)commutation relations for the elements of the smash product algebra ${\bar \cT}$
\begin{align}
 \begin{aligned}\label{smash_product_commutation_eq1}
 &\iota(\ka)\big(\ka^{(i)}\big)^{2n} = \big(\ka^{(i)}\big)^{2n} \iota(\ka), \\ &\iota(\ka)\big(\kb^{(i)}\big)^{2n} = \big(\kb^{(i)}\big)^{2n} \iota(\ka), \\ &\iota(\ka) e_+^{(i)} = e_+^{(i)} \iota(\ka), \\ &\iota(\ka)e_-^{(i)} = e_-^{(i)} \iota(\ka), \\
 \end{aligned} &&
 \begin{aligned}
 &\iota(\kb)\big(\ka^{(i)}\big)^{2n} = \big(\ka^{(i)}\big)^{2n} \iota(\kb), \\ &\iota(\kb)\big(\kb^{(i)}\big)^{2n} = \big(\kb^{(i)}\big)^{2n} \iota(\kb), \\ &\iota(\kb) e_+^{(i)} = e_+^{(i)} (\iota(\kb)+1), \\ &\iota(\kb)e_-^{(i)} = e_-^{(i)} ( \iota(\kb) - 1),
 \end{aligned}
\end{align}
and
\begin{align}
 \begin{aligned}
 &\iota(e_+)\big(\ka^{(i)}\big)^{2n} = \big(\ka^{(i)}\big)^{2n} \iota(e_+), \\ &\iota(e_+)\big(\kb^{(i)}\big)^{2n} = \big(\kb^{(i)}\big)^{2n} \iota(e_+) - q^{-n}[n]_q(q-q^{-1}) \big(\kb^{(i)}\big)^{2n} e_+^{(i)} , \\ &\iota(e_+) e_+^{(i)} = -e_+^{(i)} \iota(e_+), \\ &\iota(e_+)e_-^{(i)} = -e_-^{(i)} \iota(e_+) + \ffrac{\ka^{(i)}-\big(\ka^{(i)}\big)^{-1}}{q-q^{-1}} ,
 \end{aligned}
\end{align}
and
\begin{align}\label{smash_product_commutation_eq3}
 \begin{aligned}
 &\iota(e_-)\big(\ka^{(i)}\big)^{2n} = \big(\ka^{(i)}\big)^{2n} \iota(e_-), \\ &\iota(e_-)\big(\kb^{(i)}\big)^{2n} = \big(\kb^{(i)}\big)^{2n} \iota(e_-) + q^{n}[n]_q(q-q^{-1}) \big(\ka^{(i)}\big)^{-1} \big(\kb^{(i)}\big)^{2n} e_-^{(i)} \iota(\ka), \\ &\iota(e_-) e_+^{(i)} = -e_+^{(i)} \iota(e_-) + \ffrac{1-\big(\ka^{(i)}\big)^{-2}}{q-q^{-1}} \iota(\ka), \\ &\iota(e_-)e_-^{(i)} = -e_-^{(i)} \iota(e_-) ,
 \end{aligned}
\end{align}
for $i=a,b$. It can be checked that the above (anti-)commutation relations are reproduced from the equations \eqref{gaugerelations1}-\eqref{gaugerelations2} that are explicitly given in our case by the system of equations:
\begin{align}
 \begin{aligned}
 & (1\otimes \iota(\ka))A=A(1\otimes \iota(\ka)), \\
 & (1\otimes \iota(\kb))A=A(1\otimes \iota(\kb)), \\
 & (\ka^{-1}\otimes \iota(e_+)+e_+\otimes\iota(1))A=A(\ka^{-1}\otimes \iota(e_+)+e_+\otimes\iota(1)), \\
 & (1\otimes \iota(e_-)+e_-\otimes\iota(\ka))A=A(1\otimes \iota(e_-)+e_-\otimes\iota(\ka)) ,
 \end{aligned}
\end{align}
for the $a$-cycle, while for the $b$-cycle they are
\begin{align}
 \begin{aligned}
 & (1\otimes \iota(\ka))B=B(1\otimes \iota(\ka)), \\
 & (1\otimes \iota(\kb))B=B(1\otimes \iota(\kb)), \\
 & (\ka^{-1}\otimes \iota(e_+)+e_+\otimes\iota(1))B=B(\ka^{-1}\otimes \iota(e_+)+e_+\otimes\iota(1)), \\
 & (1\otimes \iota(e_-)+e_-\otimes\iota(\ka))B=B(1\otimes \iota(e_-)+e_-\otimes\iota(\ka)) .
 \end{aligned}
\end{align}

\subsection{The gauge-invariant subalgebra $\cA$}\label{sec:4-A}
Here, we study the gauge-invariant subalgebra $\cA$. In order to investigate it, we begin with an arbitrary vector in $\cA$ of the form
\begin{align}\label{general_gauge_invariant_element}
\begin{aligned}
f &= \sum_{\substack{n_1,m_1,\\n_2,m_2=0}}^{p-1} \sum_{\substack{r_1,s_1,\\r_2,s_2=0}}^1 f_{r_1,s_1,r_2,s_2}(n_1,m_1,n_2,m_2) \big(\ka^{(a)}\big)^{2n_1} \big(\kb^{(a)}\big)^{2m_1} \big(e_+^{(a)}\big)^{r_1 }\big(e_-^{(a)}\big)^{s_1} \times \\
& \qquad \qquad \qquad \times \big(\ka^{(b)}\big)^{2n_2} \big(\kb^{(b)}\big)^{2m_2} \big(e_+^{(b)}\big)^{r_2} \big(e_-^{(b)}\big)^{s_2 } .
\end{aligned}\end{align}
While the indices $n_1,m_1,n_2,m_2$ of the coefficients $f_{r_1,s_1,r_2,s_2}(n_1,m_1,n_2,m_2)$ are \textit{a priori} integers, we extend them to half-integers by equating the  indices $n\pm\frac{1}{2}$ with  $n+\frac{p\pm1}{2}$. We will use this convention in this and the following sections.
Then, the gauge-invariance conditions
\begin{align}\label{eq:guage-cond}
 \iota (x) f = (-1)^{|f||x|} f \iota(x), \qquad x = \ka,\kb,e_+,e_-,
\end{align}
translate to a set of conditions for the coefficients {\small
\begin{align*}
&f_{1,1,1,1}(n_1,m_1,n_2-\ffrac{1}{2},m_2) - f_{1,1,1,1}(n_1,m_1,n_2+\ffrac{1}{2},m_2) + q^{-m_2}[m_2]_q(q-q^{-1})^2 f_{1,1,0,0}(n_1,m_1,n_2,m_2)   \\
&= - q^{-m_1}[m_1]_q(q-q^{-1})^2 f_{0,1,1,0}(n_1,m_1,n_2,m_2) ,
\end{align*}
\begin{align*}
& f_{1,1,1,1}(n_1-\ffrac{1}{2},m_1,n_2,m_2) - f_{1,1,1,1}(n_1+\ffrac{1}{2},m_1,n_2,m_2) + q^{-m_1}[m_1]_q(q-q^{-1})^2 f_{0,0,1,1}(n_1,m_1,n_2,m_2)  \\
&= q^{-m_2}[m_2]_q(q-q^{-1})^2 f_{1,0,0,1}(n_1,m_1,n_2,m_2) ,
\end{align*}
\begin{align*}
&f_{1,1,0,0}(n_1-\ffrac{1}{2},m_1,n_2,m_2)
-f_{1,1,0,0}(n_1+\ffrac{1}{2},m_1,n_2,m_2)+q^{-m_1}[m_1]_q(q-q^{-1})^2 f_{0,0,0,0}(n_1,m_1,n_2,m_2)   \\
&= f_{1,0,0,1}(n_1,m_1,n_2+\ffrac{1}{2},m_2) - f_{1,0,0,1}(n_1,m_1,n_2-\ffrac{1}{2},m_2),
\end{align*}
\begin{align*}
&f_{0,0,1,1}(n_1,m_1,n_2-\ffrac{1}{2},m_2)
-f_{0,0,1,1}(n_1,m_1,n_2+\ffrac{1}{2},m_2) + q^{-m_2}[m_2]_q (q-q^{-1})^2 f_{0,0,0,0}(n_1,m_1,n_2,m_2)   \\
&= f_{0,1,1,0}(n_1-\ffrac{1}{2},m_1,n_2,m_2) - f_{0,1,1,0}(n_1+\ffrac{1}{2},m_1,n_2,m_2) ,
\end{align*}
\begin{align*}
&f_{1,1,1,1}(n_1,m_1,n_2,m_2) -
f_{1,1,1,1}(n_1+1,m_1,n_2,m_2) - q^{-m_1}[m_1]_q (q-q^{-1})^2 f_{0,0,1,1}(n_1+\ffrac{1}{2},m_1,n_2,m_2)  \\
&= - q^{m_2-2m_1} [m_2]_q (q-q^{-1})^2 f_{0,1,1,0}(n_1,m_1,n_2+\ffrac{1}{2},m_2) ,
\end{align*}
\begin{align*}
& f_{1,1,1,1}(n_1,m_1,n_2,m_2) - f_{1,1,1,1}(n_1,m_1,n_2+1,m_2)
+q^{-m_2}[m_2]_q (q-q^{-1})^2
 f_{1,1,0,0}(n_1,m_1,n_2+\ffrac{1}{2},m_2)  \\
&= q^{m_1-2m_2}[m_1]_q (q-q^{-1})^2 f_{1,0,0,1}(n_1+\ffrac{1}{2},m_1,n_2,m_2) ,
\end{align*}
\begin{align*}
& f_{0,0,1,1}(n_1,m_1,n_2,m_2) - f_{0,0,1,1}(n_1,m_1,n_2+1,m_2) + q^{-m_2} [m_2]_q (q-q^{-1})^2 f_{0,0,0,0}(n_1,m_1,n_2+\ffrac{1}{2},m_2)   \\
&= q^{2(m_1-m_2)} \left( f_{1,0,0,1}(n_1+1,m_1,n_2,m_2) - f_{1,0,0,1}(n_1,m_1,n_2,m_2) \right) ,
\end{align*}
\begin{align*}
& f_{1,1,0,0}(n_1,m_1,n_2,m_2) - f_{1,1,0,0}(n_1+1,m_1,n_2,m_2) + q^{-m_1}[m_1]_q
(q-q^{-1})^2 f_{0,0,0,0}(n_1+\ffrac{1}{2},m_1,n_2,m_2)    \\
&= q^{2(m_2-m_1)} \left( f_{0,1,1,0}(n_1,m_1,n_2,m_2)  - f_{0,1,1,0}(n_1,m_1,n_2+1,m_2) \right),
\end{align*}}
\!\!\!for $n_1,n_2,m_1,m_2=0,\ldots p-1$, and all the other coefficients are zero.
The equations above have been obtained in the following way:
 one can commute elements $\iota(x)$ through the elements from the
  expansion~\eqref{general_gauge_invariant_element}
\begin{align*}
 e_{r_1,s_1,r_2,s_2}(n_1,m_1,n_2,m_2) &= \big(\ka^{(a)}\big)^{2n_1} \big(\kb^{(a)}\big)^{2m_1} \big(e_+^{(a)}\big)^{r_1 }\big(e_-^{(a)}\big)^{s_1} \times \\
 &\qquad \times \big(\ka^{(b)}\big)^{2n_2} \big(\kb^{(b)}\big)^{2m_2} \big(e_+^{(b)}\big)^{r_2} \big(e_-^{(b)}\big)^{s_2 } ,
\end{align*}
and this produces terms coming from the non-trivial commutation relations \eqref{smash_product_commutation_eq1}-\eqref{smash_product_commutation_eq3}. Then in the sum~\eqref{general_gauge_invariant_element},  as the elements $e_{r_1,s_1,r_2,s_2}(n_1,m_1,n_2,m_2)$ are linearly independent, the coefficients in front of them have to vanish independently from one another, and this leads to a set of equations on the coefficients $ f_{r_1,s_1,r_2,s_2}(n_1,m_1,n_2,m_2)$.
First, the equations corresponding to the commutation with $\iota(\kb)$ imply that all the coefficients except $f_{1,1,1,1}$, $f_{1,1,0,0}$, $f_{0,0,1,1}$, $f_{1,0,0,1}$, $f_{0,1,1,0}$ and $f_{0,0,0,0}$ are zero. In particular, we see that  $\cA$ is  even. Then, the first four equations above were obtained from the commutation with $\iota(e_+)$, while the remaining four
were obtained in  the case with $\iota(e_-)$.

We now make an important observation:  $\cA$ contains all elements $f\in Z(\cT^{(i)})$ which are central within the $i$-cycle subalgebra $\cT^{(i)}$, or such that
\begin{align*}
 \big[f,\ka^{(i)}\big]=\big[f,\kb^{(i)}\big]=\big[f,e_\pm^{(i)}\big] = 0.
\end{align*}
 This of course follows from our general result in~\eqref{eq:Z-ab-in-A}.
 One can also see this via a direct calculation. As said above $f$ should be even, and   we further assume that $f$ can be written as a sum of products of even elements corresponding to the two cycles $a$ and $b$, and we show below that such an assumption gives non-trivial solutions. From the assumption, it follows that $r_1+s_1=0\,\mathrm{mod}\,2$ and $r_2+s_2=0\,\mathrm{mod}\,2$. In other words, we take the following Ansatz for the coefficients from~\eqref{general_gauge_invariant_element}:
 \be
 f_{r_1,s_1,r_2,s_2}(n_1,m_1,n_2,m_2) = f^{(a)}_{r_1,s_1}(n_1,m_1) f^{(b)}_{r_2,s_2}(n_2,m_2)
\ee
with
\be\label{centre_coeffs_eq1_gl11}
{f}^{(i)}_{0,1}(n,m) = {f}^{(i)}_{1,0}(n,m) = 0, \qquad  i=a,b.
\ee
So, in particular $f_{1,0,0,1}(n_1,m_1,n_2,m_2) = f_{0,1,1,0}(n_1,m_1,n_2,m_2) = 0$.
Then, the gauge-invariance equations reduce to the  following simple equation
\be\label{centre_coeffs_eq3_gl11}
  { f}^{(i)}_{1,1}(n+\ffrac{1}{2},m) - {f}^{(i)}_{1,1}(n-\ffrac{1}{2},m) = q^{-m} [m]_q (q-q^{-1})^2 {f}^{(i)}_{0,0}(n,m).
\ee

We now recall the description of the centre in Section~\ref{sec:4-centre}, and check that~\eqref{centre_coeffs_eq1_gl11} and~\eqref{centre_coeffs_eq3_gl11} are satisfied if and only if the element
$$
f^{(i)} = \sum_{n,m=0}^{p-1} \sum_{r,s=0}^1 f^{(i)}_{r,s}(n,m)\big(\ka^{(i)}\big)^{2n} \big(\kb^{(i)}\big)^{2m} \big(e_+^{(i)}\big)^{r}\big(e_-^{(i)}\big)^{s}
$$
 belongs to the centre of $\cT^{(i)}$. We establish this result using the basis of the centre provided by~\eqref{basis_of_the_centre_of_gl11_eq1}-\eqref{basis_of_the_centre_of_gl11_eq3}. We thus conlcude that the centres of the both cycle subalgebras are indeed contained in the  gauge-invariant subalgebra~$\cA$.

The subalgebra in $\cA$ generated by the two centres of $\cT^{(i)}$ turns out to be only a proper subalgebra.
Using a symbolic algebra computer program, we obtained all solutions to the gauge-invariance equations (the eight equations below~\eqref{eq:guage-cond}) for values of $p$ ranging from $3$ to $13$. Based on those results, we claim that
\be
\dim \cA = 2 p^4 + 4.
\ee
 We see that only $(p^2+1)^2$ linearly independent elements out of  $2 p^4 + 4$ are generated
 by central elements of $\cT^{(a)}$ and $\cT^{(b)}$.
  In other words, there are still many elements in $\cA$ that do not satisfy the assumption~\eqref{centre_coeffs_eq1_gl11}. For an example of such elements,
 we have
\begin{align*}
 \sum_{n,m=0}^{p-1} \big(k_\alpha^{(a)}\big)^{2n} e_\pm^{(a)} \big(k_\alpha^{(b)}\big)^{2m} e_\mp^{(b)} \; \in \; \cA,
\end{align*}
which clearly cannot be obtained as a linear combination of  products of  elements from $Z(\cT^{(a)})$ and $Z(\cT^{(b)})$. 
 We do not give a basis in $\cA$. However our aim is to describe the Fock representation of $\cA$, and for this we actually do not need to know an explicit basis -- it is turned out that only  $Z(\cT^{(a)})$  contributes to gauge-invariant states as it is explained below, of course in agreement with the general result in~\eqref{eq:RA-Z}.

\subsection{Fock representation}\label{sec:4-Fock}
Here, we investigate the Fock-type representation of $\cA$ and find an explicit basis in it.

We begin with the representation $D$ of the handle algebra $\cT$. We define the vacuum vector $|0\rangle\in\mathfrak{R}$ such that
\begin{align}
 D\big(\ka^{(b)}\big) |0\rangle =  D\big(\kb^{(b)}\big) |0\rangle = |0\rangle, && D\big(e_+^{(b)}\big) |0\rangle =  D\big(e_-^{(b)}\big) |0\rangle = 0 ,
\end{align}
recall~\eqref{eq:B-act-R}.
The vectors of the representation space
$$
\mathfrak{R}=\big\{|n,m,r,s\rangle\big\}_{n,m,r,s=0}^{p-1,p-1,1,1}
$$ are defined by the action of the $a$-cycle elements on the vacuum vector as follows
\begin{align}
 |n,m,r,s\rangle := D\Big( \big(\ka^{(a)} \big)^{2n} \big(\kb^{(a)}\big)^{2m} \big({e_+^{(a)}}\big)^r \big({e_-^{(a)}} \big)^s\Big) |0\rangle .
\end{align}
The representation $D$ extends to a representation of ${\bar \cT}$ by
\begin{align}
 D(\iota(\ka))|0\rangle = D(\iota(\kb))|0\rangle = |0\rangle, && D(\iota(e_\pm))|0\rangle = 0.
\end{align}

We now turn to the representation $\RA$ of $\cA$  defined  by~\eqref{eq:RA-def}.
The representation space $\RA$ is in fact isomorphic as a vector space to the centre $Z(\cG)$ of the Hopf algebra~$\cG$, recall our result in~\eqref{eq:RA-Z}. One can actually  check this result by a direct calculation. Indeed, using the general form \eqref{general_gauge_invariant_element} of an element in $\cA$,  we get
\begin{align*}
 D(f)|0\rangle &= \sum_{n_1,m_1=0}^{p-1} \sum_{r_1,s_1=0}^1 \left( \sum_{n_2,m_2=0}^{p-1} f_{r_1,s_1,0,0}(n_1,m_1,n_2,m_2) \right) |n_1,m_1,r_1,s_1\rangle =\\
 &=: \sum_{n_1,m_1=0}^{p-1} \sum_{r_1,s_1=0}^1 {\tilde f}_{r_1,s_1}(n_1,m_1) |n_1,m_1,r_1,s_1\rangle .
\end{align*}
One can show  that when the coefficients $f_{r_1,s_1,r_2,s_2}(n_1,m_1,n_2,m_2)$ satisfy the gauge-invariance equations -- the eight equations below~\eqref{eq:guage-cond} -- the appropriate coefficients ${\tilde f}_{r_1,s_1}(n_1,m_1)$ do satisfy the equations \eqref{centre_coeffs_eq1_gl11}-\eqref{centre_coeffs_eq3_gl11}, which are satisfied if and only if the corresponding vector belongs to the vector space
$D\big(Z(\cT^{(a)})\big)|0\rangle$. We thus conclude that the vector space of solutions is isomorphic to the centre of $\cG$.

Recall that  the centre of $\cG$ and its basis are described in Section~\ref{sec:4-centre}. The embedding map $\kappa^{(a)}$ from~\eqref{embeddings_of_G_in_T} applied to the  central elements from~\eqref{eq:central-1}-\eqref{eq:central-3}   gives the gauge-invariant vectors
\be\label{eq:def-gauge-vec}
v_{e,n}:= D\big(\kappa^{(a)}(c_{e,n})\big)|0\rangle,~~~~
x_{n}:= D\big(\kappa^{(a)}(a_{n})\big)|0\rangle,~~~~
w := D\big(\kappa^{(a)}(b)\big)|0\rangle.
\ee
From the above discussion, or from~\eqref{eq:RA-Z}, we thus have
\begin{equation}\label{eq:basis-RA}
\text{basis in}\;	\RA = \big\{\, v_{e,n}, w, x_n\,\big\}_{e=1,n=0}^{p-1,p-1} \ .
\end{equation}
A straightforward calculation gives
\begin{align}
 v_{e,n} &= (q-q^{-1})^2 [e]_q | n-\ffrac{1}{2},e,1,1\rangle - q^e\big(|n,e,0,0\rangle-|n-1,e,0,0\rangle \big) ,\\
 w &= -2(q-q^{-1})^2 \sum_{t=0}^{p-1} |t,0,1,1\rangle, \\
 x_n &= |n,0,0,0\rangle .
\end{align}
And we follow here the convention where the vectors corresponding to half-integer values are identified with those corresponding to the integer ones according to
$$
|n\pm\ffrac{1}{2},m,r,s\rangle := |n+\ffrac{p\pm1}{2},m,r,s\rangle.
$$
 Moreover, the indices $e,n$ of the elements $v_{e,n}, x_n$ are taken modulo $p$, and in what follows we use
  $$
  v_{e\pm p,n}:=v_{e,n}, \qquad v_{e,n\pm p}:=v_{e,n},  \qquad x_{n\pm p}:=x_{n}.
  $$

\subsection{$SL(2,\mathbb{Z})$ action from the handle algebra}\label{sec:handle_mcg_gl11}
In this section, we construct the realisation of the $SL(2,\mathbb{Z})$ group as operators on the representation space of the gauge-invariant subalgebra of the handle algebra. In order to do that, we use the definitions of
the mapping class group operators~\eqref{dehn_twist_quantum_eq1} corresponding to the Dehn twists along the cycles of the torus. In the end, we obtain  matrix coefficients of the $\SSS$ and $\TTT$ transformations, and check explicitly the $SL(2,\mathbb{Z})$ relations.

Proceeding as in the case of the toy model, by a direct evaluation of~\eqref{dehn_twist_quantum_eq1} we get the following expressions
\begin{equation}
 {\hat v}(i) = -\frac{i}{p} \sum_{n,m=0}^{p-1} q^{m(2n+1)} \big(\ka^{(i)}\big)^{2n} \big(\kb^{(i)}\big)^{2m} \big(1+(q-q^{-1}) \ka^{(i)} e_+^{(i)}e_-^{(i)}\big) , \qquad i=a,b.
\end{equation}
The quantum Dehn twist operators evaluated on the representation $D$ are then given by the matrix coefficients
\begin{align}\begin{aligned}
 D({\hat v}(b))|n,m,0,0\rangle &= -i q^{m(1-2n)} |n,m,0,0\rangle, \\
 D({\hat v}(b))|n,m,0,1\rangle &= -i q^{-2nm} |n,m,0,1\rangle, \\
 D({\hat v}(b))|n,m,1,0\rangle &= -i q^{m(1-2n)} |n,m,1,0\rangle, \\
 D({\hat v}(b))|n,m,1,1\rangle &= -i q^{-2nm} \left( |n,m,1,1\rangle - \ffrac{q^{2m}}{q-q^{-1}} |n-\ffrac{1}{2},m,0,0\rangle \right),
\end{aligned}\end{align}
and
\begin{align}\begin{aligned}
 D({\hat v}(a))|n,m,0,0\rangle = &-\ffrac{i}{p} \sum_{s,t=0} ^{p-1}q^{t(2s+1)} \left( |n+s,m+t,0,0\rangle + \right. \\
 &\left. + (q-q^{-1}) |n+s+\ffrac{1}{2},m+t,1,1\rangle \right), \\
 D({\hat v}(a))|n,m,0,1\rangle &= -\ffrac{i}{p} \sum_{s,t=0} ^{p-1}q^{t(2s+1)} |n+s,m+t,0,1\rangle , \\
 D({\hat v}(a))|n,m,1,0\rangle &= -\ffrac{i}{p} \sum_{s,t=0}^{p-1} q^{t(2s+1)} |n+s+1,m+t,1,0\rangle, \\
 D({\hat v}(a))|n,m,1,1\rangle &= -\ffrac{i}{p} \sum_{s,t=0} ^{p-1}q^{t(2s+1)} |n+s+1,m+t,1,1\rangle .
\end{aligned}\end{align}

\subsubsection*{$\SSS$-transformation}
On the representation $D$, the $\SSS$-transformation~\eqref{handle_alg_S_transf} is given explicitly by
\begin{align*}
 D(\SSS)|n,m,0,0\rangle &= \ffrac{i}{p} (q-q^{-1}) q^m \sum_{s,t=0}^{p-1} q^{-2nt-2ms} |s,t,1,1\rangle, \\
 D(\SSS)|n,m,0,1\rangle &= \ffrac{i}{p} q^{-m} \sum_{s,t=0}^{p-1} q^{-2(n-\ffrac{1}{2})t-2ms} |s,t,0,1\rangle , \\
 D(\SSS)|n,m,1,0\rangle &= \ffrac{i}{p} q^{2m} \sum_{s,t=0}^{p-1} q^{-2nt-2ms} |s,t,1,0\rangle, \\
 D(\SSS)|n,m,1,1\rangle &= \ffrac{i}{p} \sum_{s,t=0}^{p-1} q^{-2(n-\ffrac{1}{2})t-2ms} \left[ q^m(-1+q^{-2t}) |s,t,1,1\rangle -\ffrac{1}{q-q^{-1}} |s,t,0,0\rangle \right].
\end{align*}
Then, the action of $\SSS$ on the gauge-invariant vectors~\eqref{eq:basis-RA} in $\RA$ is
\begin{align}\label{s_transformation_handle_uqgl_eq1}
 \DA(\SSS)v_{e,n} &= \ffrac{i}{p} \sum_{\substack{s,t=0\\
 		s\neq0}}^{p-1} q^{-2s(n-\ffrac{1}{2})-2e(t-\ffrac{1}{2})} v_{s,t} - \ffrac{i}{p} (q^e-q^{-e}) \sum_{t=0}^{p-1} q^{-2et} x_t , 	\\
 \DA(\SSS)w &= 2 i (q-q^{-1}) \sum_{t=0}^{p-1} x_{t} , \\	
 \DA(\SSS)x_n &= \ffrac{i}{p (q-q^{-1})} \sum_{\substack{s,t=0\\s\neq0}}^{p-1} \ffrac{q^{-2ns}}{[s]_q} v_{s,t} - \ffrac{i}{2p (q-q^{-1})} w  .\label{s_transformation_handle_uqgl_eq3}
\end{align}
and so it can be calculated that
\begin{align*}
 &\DA(\SSS^2)v_{e,n} = - v_{-e,1-n}, && \DA(\SSS^4)v_{e,n} = v_{e,n} , \\
 &\DA(\SSS^2)w = w, && \DA(\SSS^4)w = w , \\
 &\DA(\SSS^2)x_n = x_{-n}, && \DA(\SSS^4)x_n = x_n,
\end{align*}
where $e,n$ indices are taken modulo $p$ and therefore we set $v_{-e,n}:=v_{p-e,n}$, etc. One sees that the fourth power of $\SSS$ is an identity on the subset of gauge-invariant vectors
\begin{align}\label{eq:S4-id}
 &\DA(\SSS^4) y = y, \qquad \forall y = v_{e,n},w,x_n .
\end{align}

We note the similarity between our $S$-transformation in~\eqref{s_transformation_handle_uqgl_eq1}-\eqref{s_transformation_handle_uqgl_eq3} 
and the one spelled out in~\cite[Sec.\,3.5.4]{Mikhaylov:2015qik}. Mikhaylov describes the action of the $S$-transformation on $p^2+p$ states, 
denoted by $|L_{n,m}\rangle$, $|L_m\rangle$ and $|L_{P,m}\rangle$. But these are not linearly independent. By inspection one can see that 
his representation space is spanned by $p^2+1$ linearly independent states, just as ours. More specifically, one can identify our vectors 
$v_{n,m}$, $x_m$, $w$ with the respective states $|L_{n,m}\rangle$, $|L_m\rangle$ and the sum $\sum_m |L_{P,m}\rangle$, up to some 
normalization. In this basis, Mikhaylov's action of the $S$-transformation can be seen to agree with the formulas we displayed 
above.

\subsubsection*{$\TTT$-transformation}
On the representation space $\mathfrak{R}$, $\TTT$-transformation~\eqref{handle_alg_T_transf}
  is given explicitly by
\begin{align*}
 D(\TTT)|n,m,0,0\rangle &= \ffrac{i}{p} \sum_{s,t=0}^{p-1} q^{-t(2s-1)} \Big( |n+s,m+t,0,0\rangle + \\
 &- (q-q^{-1}) |n+s-\ffrac{1}{2},m+t,1,1\rangle \Big), \\
 D(\TTT)|n,m,0,1\rangle &= \ffrac{i}{p} \sum_{s,t=0}^{p-1} q^{-t(2s-1)} |n+s,m+t,0,1\rangle , \\
 D(\TTT)|n,m,1,0\rangle &= \ffrac{i}{p} \sum_{s,t=0}^{p-1} q^{-t(2s-1)} |n+s-1,m+t,1,0\rangle, \\
 D(\TTT)|n,m,1,1\rangle &= \ffrac{i}{p} \sum_{s,t=0} ^{p-1}q^{-t(2s-1)} |n+s-1,m+t,1,1\rangle .
\end{align*}
Then, the action of $\TTT$ on gauge-invariant vectors is
\begin{align}
 \DA(\TTT) v_{e,n} &= \ffrac{i}{p} \sum_{\substack{s,t=0,\\ s\neq -e}}^{p-1} q^{-2st} v_{e+s,n+t} +\ffrac{i}{p} (q^e-q^{-e}) \sum_{t=0}^{p-1} q^{2e(t-n+\ffrac{1}{2})} x_t ,\\
 \DA(\TTT) w &= i w , \\
 \DA(\TTT) x_n &= -\ffrac{i}{p (q-q^{-1})} \sum_{\substack{s,t=0 \\ s\neq 0}}^{p-1} \ffrac{q^{-s(2t-1)}}{[s]_q} v_{s,n+t} + \ffrac{i}{2p  (q-q^{-1})} w + \ffrac{i}{p} \sum_{t=0}^{p-1} x_{t} .
\end{align}
It can be calculated that in fact, together with the $\SSS$-transformation, the $\TTT$-transformation defined in this way provides an action of $SL(2,\mathbb{Z})$ on 
the sub-space of gauge-invariant vectors
\begin{align}
 &\DA\big((\SSS\TTT)^3\big) y = \DA(\SSS^2) y, \qquad \forall y = v_{e,n},w,x_n .
\end{align}

Let us once again compare with the formulas for the $T$-transformation in \cite{Mikhaylov:2015qik}. At first sight the 
two sets of formulas look quite different even after Mikhaylov's formulas are written in a proper basis, as we described 
after~\eqref{eq:S4-id}. But the two representations turn out to be equivalent via the conjugation with the modular $S$-matrix. Put differently, our $T$-transformation was defined through a Dehn 
twist along the $a$-cycle. But equivalently, one can also use the Dehn twist along the $b$-cycle. These two Dehn twists 
are related by a modular $S$-transformation and, in our terminology, it is the Dehn twist along the $b$-cycle that is 
described by the formulas in~\cite{Mikhaylov:2015qik}. In conclusion, the representation of the modular group we 
obtained through the general formalism we developed in Section~\ref{Sec:2} is equivalent to the one of Mikhaylov 
when the gauge group $G$ of our Chern-Simons theory is $G=\GL$.

\subsection{Lyubashenko--Majid $SL(2,\mathbb{Z})$ action on the centre}\label{sec:LM_mcg_gl11}
In this section, we consider the $SL(2,\mathbb{Z})$ action of LM type and compute the $\SSS$- and $\TTT$-transformations on the basis elements of $Z(\UBglsimple)$ constructed in Section~\ref{sec:4-centre}. Moreover, we check that these transformations  provide a  projective $SL(2,\mathbb{Z})$ action indeed.

\subsubsection*{$\SSS$-transformation}

Using equation \eqref{eq:S-transf}, we find  the LM-type  $\SSS$-transformation
\begin{align}
 \SSSZ(c_{e,n}) &= -\ffrac{i}{p} \sum_{\substack{s,t=0\\ s\neq0}}^{p-1} q^{-2e(t-\ffrac{1}{2})-2s(n-\ffrac{1}{2})} c_{s,t}
 +\ffrac{i}{p} (q^e-q^{-e}) \sum_{t=0}^{p-1} q^{-2et} a_{t} , \label{eq:S-LM-1}
 \\
 \SSSZ(b) &= -2 i (q-q^{-1}) \sum_{t=0}^{p-1} a_{t} , \\
 \SSSZ(a_n) &= -\ffrac{i}{p (q-q^{-1})} \sum_{\substack{s,t=0, \\ s\neq0}}^{p-1} \ffrac{q^{-2ns}}{[s]_q} c_{s,t} + \ffrac{i}{2 p (q-q^{-1})} b .\label{eq:S-LM-3}
\end{align}
Iterative application of the formulae~\eqref{eq:S-LM-1}-\eqref{eq:S-LM-3} then leads to
\begin{align*}
 &\SSSZ^2(c_{e,n}) = - c_{-e,-n+1}, &&\SSSZ^4(c_{e,n})  = c_{e,n} ,\\
 &\SSSZ^2(b) =  b, && \SSSZ^4(b)  = b, \\
 &\SSSZ^2(a_n) =  a_{-n}, && \SSSZ^4(a_n) = a_n,
\end{align*}
where we recall that $e,n$ indices are taken modulo $p$ and so we have $c_{-e,n}:=c_{p-e,n}$, etc.
We thus see that the fourth power of the $\SSS$-transformation is an identity on the centre of $\UBglsimple$:
\begin{align}
 \SSSZ^4 = id .
\end{align}
\subsubsection*{$\TTT$-transformation}
Using equation \eqref{eq:T-transf} we find the action of the $\TTT$-transformation on the central elements of $\UBglsimple$ 
\begin{align}
 \TTTZ(c_{e,n}) &= \ffrac{1}{p} \sum_{\substack{s,t=0,\\ s\neq -e}}^{p-1} q^{-2st} c_{e+s,n+t} + \ffrac{1}{p} \sum_{t=0}^{p-1} q^{2e(t+\ffrac{1}{2})} (q^e-q^{-e})  a_{n+t},\\
 \TTTZ(b) &= b ,\\
 \TTTZ(a_n) &= -\ffrac{1}{p (q-q^{-1})} \sum_{\substack{s,t=0,\\ s\neq 0}}^{p-1} \ffrac{q^{-s(2t-1)}}{[s]_q} c_{s,n+t} + \ffrac{1}{p} \sum_{t=0}^{p-1} a_{t} + \ffrac{1}{2p(q-q^{-1})} b .
\end{align}
Iteratively applying the above $\TTT$-transformation, as well as the $\SSS$-transformation considered above, we can establish that on the centre of $\UBglsimple$ we have the following identity
\begin{align} \label{center-relation}
 &(\SSSZ\TTTZ)^3= -i \SSSZ^2.
\end{align}
Therefore, we see that on the centre of the quantum group we have a projective $SL(2,\mathbb{Z})$ action.
\subsection{Equivalence of two actions}\label{sec:4-comparison}
In this section, we show that the two $SL(2,\bZ)$ actions presented in the sections above agree projectively.

We first note that  the centre $Z(\UBglsimple)$ and the representation space $\RA$ of the gauge-invariant subalgebra $\cA$ are isomorphic as vector spaces, in agreement with~\eqref{eq:RA-Z}.
Explicitly, we have the correspondence, recall definitions in~\eqref{eq:def-gauge-vec}:
\begin{align}
Z(\UBglsimple) \ni c_{e,n} &\;\stackrel{}{\longmapsto}\; v_{e,n} \in \RA , \\
Z(\UBglsimple) \ni a_{n} &\;\stackrel{}{\longmapsto}\; x_{n} \in \RA , \\ Z(\UBglsimple) \ni b &\;\stackrel{}{\longmapsto}\; w \in \RA .
\end{align}
Moreover, if we take into account the above isomorphism between  $Z(\UBglsimple)$ and $\RA$, we can compare the coefficients of the relevant actions in the two cases. In order to do that, let us define the coefficients of the $\SSS$-action for the handle algebra:
\begin{align*}
 \DA(\SSS)v_{n,m} &= \sum_{\substack{s,t=0\\s\neq0}}^{p-1} (\SSS_\mathcal{T})^{s,t}_{n,m} v_{s,t} + \sum_{t=0}^{p-1} (\SSS_\mathcal{T})^{t}_{n,m} x_{t} + (\SSS_\mathcal{T})^{\bullet}_{n,m} w, \\
 \DA(\SSS)x_{m} &= \sum_{\substack{s,t=0\\s\neq0}}^{p-1} (\SSS_\mathcal{T})^{s,t}_{m} v_{s,t} + \sum_{t=0}^{p-1} (\SSS_\mathcal{T})^{t}_{m} x_{t} + (\SSS_\mathcal{T})^{\bullet}_{m} w, \\
 \DA(\SSS)w &= \sum_{\substack{s,t=0\\s\neq0}}^{p-1} (\SSS_\mathcal{T})^{s,t}_{\bullet} v_{s,t} + \sum_{t=0}^{p-1} (\SSS_\mathcal{T})^{t}_{\bullet} x_{t} + (\SSS_\mathcal{T})^{\bullet}_{\bullet} w,
\end{align*}
and for the  $\TTT$-action as
\begin{align*}
 \DA(\TTT)v_{n,m} &= \sum_{\substack{s,t=0\\s\neq0}}^{p-1} (\TTT_\mathcal{T})^{s,t}_{n,m} v_{s,t} + \sum_{t=0}^{p-1} (\TTT_\mathcal{T})^{t}_{n,m} x_{t} + (\TTT_\mathcal{T})^{\bullet}_{n,m} w, \\
 \DA(\TTT)x_{m} &= \sum_{\substack{s,t=0\\s\neq0}}^{p-1} (\TTT_\mathcal{T})^{s,t}_{m} v_{s,t} + \sum_{t=0}^{p-1} (\TTT_\mathcal{T})^{t}_{m} x_{t} + (\TTT_\mathcal{T})^{\bullet}_{m} w, \\
 \DA(\TTT)w &= \sum_{\substack{s,t=0\\s\neq0}}^{p-1} (\TTT_\mathcal{T})^{s,t}_{\bullet} v_{s,t} + \sum_{t=0}^{p-1} (\TTT_\mathcal{T})^{t}_{\bullet} x_{t} + (\TTT_\mathcal{T})^{\bullet}_{\bullet} w.
\end{align*}
 And similarly for the transformations on $Z(\cG)$:
\begin{align*}
 \SSSZ(c_{n,m}) &= \sum_{\substack{s,t=0\\s\neq0}}^{p-1} (\SSSZ)^{s,t}_{n,m} c_{s,t} + \sum_{t=0}^{p-1} (\SSSZ)^{t}_{n,m} a_{t} + (\SSSZ)^{\bullet}_{n,m} b, \\
 \SSSZ(a_{m}) &= \sum_{\substack{s,t=0\\s\neq0}}^{p-1} (\SSSZ)^{s,t}_{m} c_{s,t} + \sum_{t=0}^{p-1} (\SSSZ)^{t}_{m} a_{t} + (\SSSZ)^{\bullet}_{m} b, \\
 \SSSZ(b) &= \sum_{\substack{s,t=0\\s\neq0}}^{p-1} (\SSSZ)^{s,t}_{\bullet} c_{s,t} + \sum_{t=0}^{p-1} (\SSSZ)^{t}_{\bullet} a_{t} + (\SSSZ)^{\bullet}_{\bullet} b,
 \end{align*}
 and
 \begin{align*}
 \TTTZ(c_{n,m}) &= \sum_{\substack{s,t=0\\s\neq0}}^{p-1} (\TTTZ)^{s,t}_{n,m} c_{s,t} + \sum_{t=0}^{p-1} (\TTTZ)^{t}_{n,m} a_{t} + (\TTTZ)^{\bullet}_{n,m} b, \\
 \TTTZ(a_{m}) &= \sum_{\substack{s,t=0\\s\neq0}}^{p-1} (\TTTZ)^{s,t}_{m} c_{s,t} + \sum_{t=0}^{p-1} (\TTTZ)^{t}_{m} a_{t} + (\TTTZ)^{\bullet}_{m} b, \\
 \TTTZ(b) &= \sum_{\substack{s,t=0\\s\neq0}}^{p-1} (\TTTZ)^{s,t}_{\bullet} c_{s,t} + \sum_{t=0}^{p-1} (\TTTZ)^{t}_{\bullet} a_{t} + (\TTTZ)^{\bullet}_{\bullet} b.
\end{align*}

 It can be read-off that the non zero coefficients are
\begin{align*}
&(\SSS_\mathcal{T})^{s,t}_{n,m} = - (\SSSZ)^{s,t}_{n,m} = \ffrac{i}{p} q^{-2s(m-\ffrac{1}{2})-2n(t-\ffrac{1}{2})},
&&(\TTT_\mathcal{T})^{s,t}_{n,m} = i (\TTTZ)^{s,t}_{n,m} = \ffrac{i}{p} q^{-2(s-n)(t-m)},\\
&(\SSS_\mathcal{T})^{t}_{n,m} = - (\SSSZ)^{t}_{n,m} = -\ffrac{i}{p} (q^n-q^{-n}) q^{-2nt},
&&(\TTT_\mathcal{T})^{t}_{n,m} = i (\TTTZ)^{t}_{n,m} = \ffrac{i}{p} (q^n-q^{-n}) q^{2n(t-m+\ffrac{1}{2})} ,\\
&(\SSS_\mathcal{T})^{s,t}_{m} = - (\SSSZ)^{s,t}_{m} =\ffrac{i}{p(q-q^{-1})} \ffrac{q^{-2ms}}{[s]_q},
&&(\TTT_\mathcal{T})^{s,t}_{m} = i (\TTTZ)^{s,t}_{m}
= -\ffrac{i}{p(q-q^{-1}) } \ffrac{q^{-2s(t-m-\ffrac{1}{2})}}{[s]_q},
\\
&(\SSS_\mathcal{T})^{\bullet}_{m} = - (\SSSZ)^{\bullet}_{m} =
-\ffrac{i}{2p(q-q^{-1})}
,
&&(\TTT_\mathcal{T})^{t}_{m} = i (\TTTZ)^{t}_{m} = \ffrac{i}{p},\\
&(\SSS_\mathcal{T})^{t}_{\bullet} = - (\SSSZ)^{t}_{\bullet} = 2i(q-q^{-1}),
&&(\TTT_\mathcal{T})^{\bullet}_{m} = i (\TTTZ)^{\bullet}_{m} =\ffrac{i}{2p(q-q^{-1})}
,\\
&
&&(\TTT_\mathcal{T})^{\bullet}_{\bullet} = i (\TTTZ)^{\bullet}_{\bullet} = i .
\end{align*}
Comparing the coefficients of $\SSS$ and $\TTT$ from the handle algebra to the ones $\SSSZ$ and $\TTTZ$ from the LM construction, we see that they indeed agree projectively:
\begin{align}
 &\SSSZ = -\SSS_{\mathcal{T}},
 &&\TTTZ = -i \TTT_{\mathcal{T}} .
\end{align}

\section{Outlook}

In this work, we considered the quantisation of $\GL$ Chern-Simons theory at odd integer level
on a torus $\Sigma = \Sigma_{1,0} = \mathbb{T}^2$ with no punctures. While the general framework of
combinatorial quantisation allows to consider  an arbitrary simplicial decomposition of $\Sigma$, we only
considered the minimal decomposition of the torus with a single 2-cell, two 1-cells and one 0-cell.
There are a number of extensions that we shall address in forthcoming work.

To begin with, we will replace the torus $\mathbb{T}^2$ by a Riemann surface $\Sigma= \Sigma_{g,n}$ of
arbitrary genus $g$ and with any number $n$ of punctures. The first step is then to choose some
simplicial decomposition. The minimal choice would involve $(n+1)$ number of 2-cells, $(2g+n)$ 1-cells and a
single 0-cell. If we adopt this choice, the monodromy (or loop) algebra
 we have discussed in this work is the only building block that is used in the combinatorial quantisation. Of course, one
needs as many of these algebras as there are 1-cells and they satisfy exchange relations that
must reflect the topology of our surface, generalising what we saw here for the torus. For more
general simiplicial decomposition with more than one 0-cell, one needs a second building block,
the holonomy (or link) algebra. It is a close relative of the $\GL$ quantum group, i.e.\ of the
Hopf-dual for $\GL$. Once introduced, link and loop algebras must be combined into a larger
algebraic structure in which they satisfy a system of exchange relations which are determined
by the simplicial decomposition and by the $R$-matrix of  $\cG$. All this will be discussed in detail in forthcoming work. There
we will also show that the spaces of Chern-Simons states are actually independent of the
simplicial decomposition so that the minimal choice can always be adopted.

The construction of representations of the modular group $SL(2,\mathbb{Z})$ that was our main focus
above  also possesses a natural extension to $\Sigma = \Sigma_{g,n}$.
In fact, for higher genus and in the presence of punctures, the modular group gets replaced by the (pure) mapping class group
of the $n$-punctured surface. The fundamental generators are the Dehn twists along non-contractible curves on $\Sigma$. To construct
representations of the mapping class group, we can follow precisely the constructions we have described in this work. 
All we need to prescribe are the corresponding Chern-Simons observables that
are associated with the non-contractible curves on $\Sigma$. In this step we can use the
same formula as for the two non-trivial cycles of the torus, see \eqref{dehn_twist_quantum_eq1}.
In some sense, one key result of the present work was to show that this prescription is equivalent
to Lyubachenko--Majid's construction for the torus as well as to Mikhaylov's representation~\cite{Mikhaylov:2015qik}  of the 
modular group,
at least for $\GL$ at integer level. Once this is established, the inherent factorisability of the 
combinatorial prescription provides a canonical extension to punctured surfaces of higher genus. 
Constructing the corresponding representation of the
mapping class group is one of the main goals of our future work. Again, our construction will be
restricted to $\GL$ Chern-Simons theory at integer level.

There are two additional extensions we are planning to describe in forthcoming papers. One
of them is to go beyond the case of integer levels. In other words, we want to admit deformation
parameters $q$ which are no longer given by a root of unity. Very little is known from other
approaches about such an extension. So, it seems worthwhile to look at it in the case of the
torus $\Sigma = \mathbb{T}^2$ first. Once the theory for the torus is developed, the combinatorial
approach provides a straightforward extension to other surfaces.

The final step is then to go beyond $\GL$. As we have mentioned in the introduction, for
2-dimensional supergroup WZNW models the quantisation resembles that of the
$\GL$ model whenever the gauge group $G$ is of type I. Given the usual duality between
WZW models and Chern-Simons theory, we expect the same to be true for the 3-dimensional
model. One of the more immediate goals therefore is to develop the combinatorial
quantisation of supergroup Chern-Simons theory for gauge supergroups $G$ of type I,
at least as long as the level is integer. 
The integer level is important here as it reduces the quantum symmetry to a \textsl{finite-dimensional} super Hopf algebra -- the case where our general construction in Section~\ref{Sec:2} is applicable. Carrying out these extensions, we hope to
construct a plethora of new representations of mapping class groups for 2-dimensional
surfaces $\Sigma_{g,n}$ or arbitrary genus $g$.

\appendix

%%%%%%%%%%%%%%%%%%%%%%%%%%%%%%%%%
\section*{Acknowledgments}
We thank Hubert Saleur for stimulating discussions.
The work of N.A. was supported by the Swiss National Science Foundation (pp00p2-157571/1). N.A., A.M.G. and M.P. are very grateful to DESY for a generous support during 2016 when this work actually started.  The work of A.M.G. was supported by CNRS.  The work of M.P. was supported by the European Research Council (advanced grant NuQFT). M.P. and N.A. acknowledges the AEC centre at University of Bern and IPhT CEA Saclay for their hospitality in the final stage of this project.
A.M.G. is also grateful to IPHT Saclay for kind hospitality in 2017 and 2018.

\providecommand{\href}[2]{#2}\begingroup\raggedright

\endgroup

\end{document}